\documentclass{article} 

\usepackage{graphicx}
\usepackage{dcolumn}
\usepackage{bm}

\usepackage{amsmath,amsfonts}
\graphicspath{{./figs/}}

\usepackage{mathtools,xparse}
\DeclarePairedDelimiter{\norm}{\lVert}{\rVert}
\NewDocumentCommand{\normL}{ s O{} m }{%
  \IfBooleanTF{#1}{\norm*{#3}}{\norm[#2]{#3}}_{L_2(D)}%
}

\newcommand{\pd}[2]{\frac{\partial#1}{\partial#2}}
\newcommand{\E}{\mathrm{E}}

\newcommand{\divU}{\nabla \cdot u}

\begin{document}
\title{Adjoint-based trailing edge shape optimization of a transonic turbine vane using
large eddy simulations}

\author{Chaitanya Talnikar$^1$ \and Qiqi Wang$^1$}
\date{%
    $^1$Massachusetts Institute of Technology\\%
    \today
}


\maketitle    

\begin{abstract}
    The shape of the  trailing edge of a gas turbine nozzle guide vane has
    a significant effect on the downstream
    stagnation pressure loss and heat transfer over the surface of the vane. 
    Traditionally, adjoint-based design optimization methods
    for turbomachinery components have used low-fidelity simulations
    like Reynolds averaged Navier-Stokes. 
    To reliably capture the complex flow phenomena
    involved in turbulent flow over a turbine vane,
    high-fidelity simulations like large eddy simulation (LES) are required.
    In this paper, an adjoint-based trailing edge shape optimization using LES is performed
    to reduce pressure loss and heat transfer over the surface of the vane.
    The chaotic dynamics of turbulence limits the effectiveness of the adjoint
    method for long-time averaged objective functions computed from LES.
    A viscosity stabilized unsteady adjoint method is used to 
    obtain gradients of the design objective function with reasonable accuracy.
    A gradient utilizing Bayesian optimization is used to robustly handle noise in the
    objective function and gradient evaluations.
    The trailing edge shape is parameterized using a linear combination of $5$ convex
    designs. 
    Results from the optimization, performed on the 
    supercomputer Mira, are compared with optimal designs
    generated using derivative-free
    design optimization of the same problem.
\end{abstract}




\section{INTRODUCTION}
In a gas turbine engine the nozzle guide vanes are responsible
for directing air onto the turbine blades. Shape design
of these vanes is an important problem as it influences
the amount of work that can be extracted from the gases exiting
the upstream combustion chamber \cite{greitzer2007internal}. In particular, the downstream
stagnation pressure loss is quite sensitive to the shape of the
trailing edge of the vane. It can alter
the boundary layer development process and move the location
of the separation points on the surface of the vane \cite{talnikar2014parallel}. 
Additionally, the trailing edge shape impacts
the amount of heat transfer from the hot gas to the 
surface of the vane. Reducing the heat transfer can significantly
increase the life span of the turbine vane \cite{han2012gas}.

Adjoint-based design optimization is an optimization tool used by 
researchers in many computational fluid dynamics (CFD)
applications for improving the performance of various
fluid machinery components. It uses the adjoint method to obtain
gradients of the design objective with respect to design parameters \cite{pironneau1974optimum,giles2000introduction}.
Gradient-based optimization algorithms require fewer iterations
to reach locally optimal designs and can scale to 
a larger number of design parameters than the typical
derivative-free optimization algorithms used in CFD \cite{marsden2007trailing,Forrester2177,CAMPANA2006634}.
In the field of aerospace engineering,
Jameson \cite{jameson1995optimum} performed
an adjoint-based shape optimization of an airfoil using the steady state Euler equations and
Lyu \cite{economon2013viscous}
performed a design optimization of a blended wing body aircraft using the Reynolds Averaged Navier-Stokes (RANS) equations.
Majority of the adjoint-based design optimizations carried out in literature
use low-fidelity simulations like RANS or Euler \cite{pironneau1992optimal,jameson1995optimum,economon2013viscous,lyu2014aerodynamic}.
To accurately model the complex fluid flow structures in the 3-dimensional turbine vane problem,
like flow separation, turbulent boundary layer development and wake mixing, high
fidelity simulations like Large Eddy Simulations (LES) are required \cite{morata2012effects,gourdain2012comparison,medic2012prediction,bhaskaran2011heat,kawai2010large,laskowski2016}.
Gourdain \cite{gourdain2012comparison}
compared LES to RANS on a similar turbine vane, 
and found that LES predicts heat transfer with a much higher accuracy.
Coupling the accuracy of LES with the performance of the current generation of supercomputers, 
adjoint-based design using LES is becoming a more practical option \cite{moin1997tackling}.

The application of the adjoint method to LES introduces certain
complications. The magnitude of the adjoint solution field diverges
to infinity as the LES is performed for a longer time \cite{wang2013drag,blonigan2012towards,bloniganpatrick,nisha2019,angxiu2019,angxiu2020}.
This divergence introduces a significant error in the gradient computed from the adjoint solution, especially
when the design objective function is a long time-averaged quantity.
The authors of this paper have proposed a fix to the problem of diverging adjoint solutions
by modifying the adjoint method for unsteady compressible Navier-Stokes equations
\cite{talnikar2016}.
Minimal artificial viscosity is added to the equations governing
the adjoint solution in regions of the flow where the magnitude of the adjoint solution is
believed to be growing exponentially fast. 
The regions of divergence of the adjoint solution are found
by performing an $L_2$ norm analysis of the adjoint equations.
Using the modified adjoint equations, the divergence of the adjoint solutions is restricted,
while maintaining a reasonable accuracy of the computed gradients.

The noise in the design objective and gradient evaluations 
and the prohibitive computational cost of the evaluations makes design optimization
with LES a challenging task \cite{talnikar2015optimization}. Numerous optimizations algorithms have been designed for handling
noisy and expensive objective functions \cite{rios2013derivative}. 
One of the most basic methods is the Robbins-Monro algorithm \cite{robbins1951stochastic}, which
decides the next design point to evaluate by taking a variable step in the gradient
direction, similar to steepest descent. The algorithm 
has the tendency to get stuck in local optimums and requires extensive tuning of the 
step parameter \cite{spall1992multivariate,polyak1992acceleration}.
Another optimization algorithm is SNOBFIT (Stable noisy branch and fit) \cite{huyer2008snobfit}, which belongs to the 
divide and conquer class of optimization algorithms. While
the algorithm is robust to noise, it cannot effectively utilize gradient
information from all past evaluations.
Bayesian optimization algorithms have a lot of properties that make them ideal
for design optimization using LES \cite{jones1998efficient,wu2017bayesian}. 
In this paper, the standard Bayesian optimization algorithm is 
modified to make it work better for noisy evaluations. 

The adjoint-based design optimization tool,
using the artificial viscosity based adjoint method and the Bayesian optimization algorithm,
is applied to design the trailing edge shape of a gas turbine nozzle guide vane.
The design objective is a weighted combination of the 
time and mass-flow averaged stagnation pressure loss coefficient
downstream of the vane and the time-averaged Nusselt number over the trailing edge surface of the vane. 
The shape of the trailing edge is parameterized using a linear combination of
$5$ convex designs (convex about the chord line). This makes the dimension
of the optimization problem equal to $4$. The optimization is performed on the
Argonne National Lab Mira supercomputer and a small-scale CPU and GPU based compute cluster. 
Section \ref{s:problem_setup} describes the physics, numerical approximation and 
parameterization procedure of the trailing edge shape for the 
flow over the 3-dimensional turbine vane. 
Section \ref{s:adjoint} details the artificial viscosity based adjoint method,
the modified Bayesian optimization algorithm and it's application
to a 2-dimensional optimization problem with the Rastigrin function
as the objective function.
Finally, section \ref{s:results} discusses
results from the adjoint-based design optimization and demonstrates
a comparison between gradient-based and derivative-free Bayesian optimization.
Section \ref{s:conclusion} concludes the paper and examines avenues for future research and 
application to new design problems.

\section{Flow over turbine vane}
\label{s:problem_setup}
The shape of a nozzle guide vane in a gas turbine engine plays
a significant role in the work extraction efficiency of the turbine.
Lower work extraction from the fluid results from a higher pressure loss 
due to an unfavorable shape of the vane.
Additionally, the shape impacts the amount of heat transfer from the hot gas to the vane 
surface, hence, determining the cooling fluid requirements for the vane \cite{han2012gas}. 
An optimally designed vane can lead to notable fuel and
repair cost savings and increased longevity of the vane.
The reference (baseline) shape
for the nozzle guide vane used for the shape optimization is
designed by researchers at the Von Karman Institute (VKI) \cite{arts1992aero}.

\subsection{Flow physics}
Understanding the physics of the flow over the turbine vane is
crucial for constructing the optimization problem.

In a gas turbine engine, gases exiting from the combustion chamber 
impinge on a circular cascade of nozzle guide vanes. 
High temperature and high pressure subsonic flow hits the leading edge of each vane
and then accelerates over the suction side, reaching
close to Mach one near the trailing edge on the suction side of the vane where the boundary layer 
transitions from laminar to turbulent. In contrast, on the 
pressure side, the boundary layer stays laminar.
The boundary layers after separation from the surface of the vane,
merge into a turbulent wake. The exit pressure in gas turbine engines is
much larger than $1$ atm.

In the experimental setup of the VKI vane, the circular cascade 
is approximated as a linear cascade.
The Reynolds number of the flow, computed using the chord length of
the vane, is approximately 1,000,000. The downstream isentropic Mach number is $0.92$.
The chord length of the vane is $c_l = 67.6\,mm$.
The Reynolds number 
is defined using the chord length of
the vane and the density, velocity and viscosity 
of the inflow.
Gourdain \cite{gourdain2012comparison} and Morata \cite{morata2012effects}
performed large eddy simulations of this flow problem at various 
Reynolds numbers.

The boundary layer development on the two sides of
the vane and turbulent mixing in the wake lead to
substantial drop in the stagnation pressure of the fluid.
These phenomena discourage the use of blunt trailing edge for the vane
as they can lead to earlier separation and higher pressure loss.
There is a large
increase in the heat transfer coefficient on the suction side
due to the transition of the boundary layer from laminar to turbulent.
This behavior rules out the use of sharp trailing edge for the vane
as the material of the vane cannot simultaneously sustain high temperatures 
and high stress for prolonged time periods.
The thickness, development and separation location of the boundary layers
and the characteristics of the turbulent wake 
are greatly influenced by the shape of the 
vane near the trailing edge. Hence, the design optimization restricts
shape parameterization of the vane to the trailing edge in order to
lower the dimension of the design search space and maintain the 
enhancement potential of the candidate designs.

\subsection{Numerical setup}
Turbulent fluid flow is modeled using the 
compressible Navier-Stokes equations with the ideal
gas law serving as an approximation to the thermodynamic state equation \cite{garnier2009large}.
The gas is assumed to be air.
\begin{align}
\begin{split}
    \mathrm{In} \hspace{2mm} &\mathbf{x} \in V, t \in [0, T],\\
    &\frac{\partial \rho}{\partial t} + \nabla \cdot (\rho \mathbf{u}) = s_\rho \\
    &\frac{\partial (\rho \mathbf{u})}{\partial t} + \nabla \cdot (\rho \mathbf{u}\mathbf{u}) + \nabla p = \nabla \cdot \sigma + \mathbf{s}_{\rho \mathbf{u}}\\
    &\frac{\partial (\rho E)}{\partial t} + \nabla \cdot (\rho E \mathbf{u} + p \mathbf{u}) = \nabla \cdot (\mathbf{u} \cdot \sigma + \alpha \rho \gamma \nabla e) + s_{\rho E} \\
    &\sigma = \mu (\nabla \mathbf{u} + \nabla \mathbf{u}^T) - \frac{2\mu}{3}(\nabla \cdot \mathbf{u})\mathbf{I} \\
    &p = (\gamma - 1)\rho e \\
    &e = E - \frac{\mathbf{u}\cdot \mathbf{u}}{2} \\
    &c = \sqrt{\frac{\gamma p}{\rho}}
\end{split}
\label{e:navier}
\end{align}
where $V$ is the domain of the flow problem,
$T$ is the terminal time
of the flow problem (which can be $\infty$), $\rho$ is the density, $\mathbf{u}$ is the velocity vector, $\rho E$ is the total energy,
$p$ is pressure, $e$ is internal energy of the fluid, $c$ is
the speed of sound,
$\gamma$ is the isentropic expansion factor (for air $\gamma = 1.4$),
$\mu$ is the viscosity field modeled
using Sutherland's law for air 
\begin{equation}
    \mu = \frac{C_s T^{3/2}}{T + T_s}
\end{equation}
where $T_s = 110.4\,K$ and $C_s = 1.458 \times 10^{-6} \frac{kg}{m\,s\,\sqrt{K}}$.
$\alpha$ is the thermal diffusivity modeled using
\begin{equation}
    \alpha = \frac{\mu}{\rho Pr}
\end{equation}
where $Pr$ is the Prandtl number (for air $Pr = 0.71$).
$s_\rho, \mathbf{s}_{\rho\mathbf{u}}, s_{\rho E}$ denote the source terms prescribed in a
flow problem.
In addition to the above equations, depending
on the specifics of the flow problem, boundary conditions 
are prescribed on each of the boundary regions of the domain ($S$). Lastly,
appropriate initial conditions for the flow variables, $\rho, \rho \mathbf{u}, \rho E$, are
defined.
the following reference quantities are defined for the flow problem
\begin{align}
    \begin{split}
    \rho_r &= 1\,kg/m^3 \\
    u_r &= 100\,m/s \\
        p_r &= C_r\rho_r u_r^2  \\
        \mu_r &= 1.8\times10^{-5} \,m^2/s
    \end{split}
\end{align}
where $C_r = 10.1325$ is a fixed constant.

The numerical approximation to the flow solution
is obtained using large eddy simulations on a discretized
domain of the 3-dimensional flow problem,
generally known as a mesh or a grid.
In an LES, the large scale eddies
of the flow are resolved by the grid while 
the contribution of the small scale eddies to the filtered 
Navier-Stokes equations are modeled using a sub-grid scale Reynolds stress model \cite{garnier2009large}.
The choice of the LES model can have a large impact on the accuracy of the relevant statistical 
quantities of interest of the flow.
In this paper, implicit LES model is used.
In this model, the numerical error of the discretization scheme serves as the LES model.
It has been shown that when using a dissipative discretization method,
the numerical viscosity from the grid can
be of the same order of magnitude as the sub-grid scale viscosity \cite{moeng1989evaluation}.
Hence, using an explicit LES model may be unnecessary.

The numerical solution of the flow problem is obtained on an
unstructured hexahedral mesh using a
second order finite volume method (FVM) \cite{leveque2002finite}.
The central differencing scheme is used to 
interpolate cell averages of the flow solution onto faces
of the mesh \cite{versteeg1995computational}. The numerical fluxes for the conservative 
flow variables are computed using
the Roe approximate Riemann solver \cite{roe1981approximate}. 
An explicit time integration scheme, the
strong stability preserving third order Runge-Kutta method \cite{macdonald2003constructing}, is used
for time marching the numerical flow solution.
The size of the time step is determined using the acoustic
Courant-Friedrichs-Lewy (CFL) condition \cite{courant1967partial}.
The flow solver is implemented in Python using 
a library, \textit{adFVM}\cite{talnikarhpc}, that provides a high-level abstract application
programming interface for writing efficient CFD applications.
The flow solver is parallelized using the Message Passing
Interface (MPI) library. 
The adjoint flow solver is implemented as a discrete 
adjoint solver using the Python
library \textit{adFVM} with the help
of automatic differentiation. 
The checkpointing method \cite{wang2009minimal} is used to reduce memory usage.
The viscosity-stabilized adjoint method, described in Section
\ref{s:adjoint}, is implemented for
the adjoint flow solver
using an Euler-RK implicit-explicit time stepping scheme \cite{ascher1997implicit}.

\begin{figure}[htb!]
\begin{center}
    \includegraphics[width=0.8\textwidth]{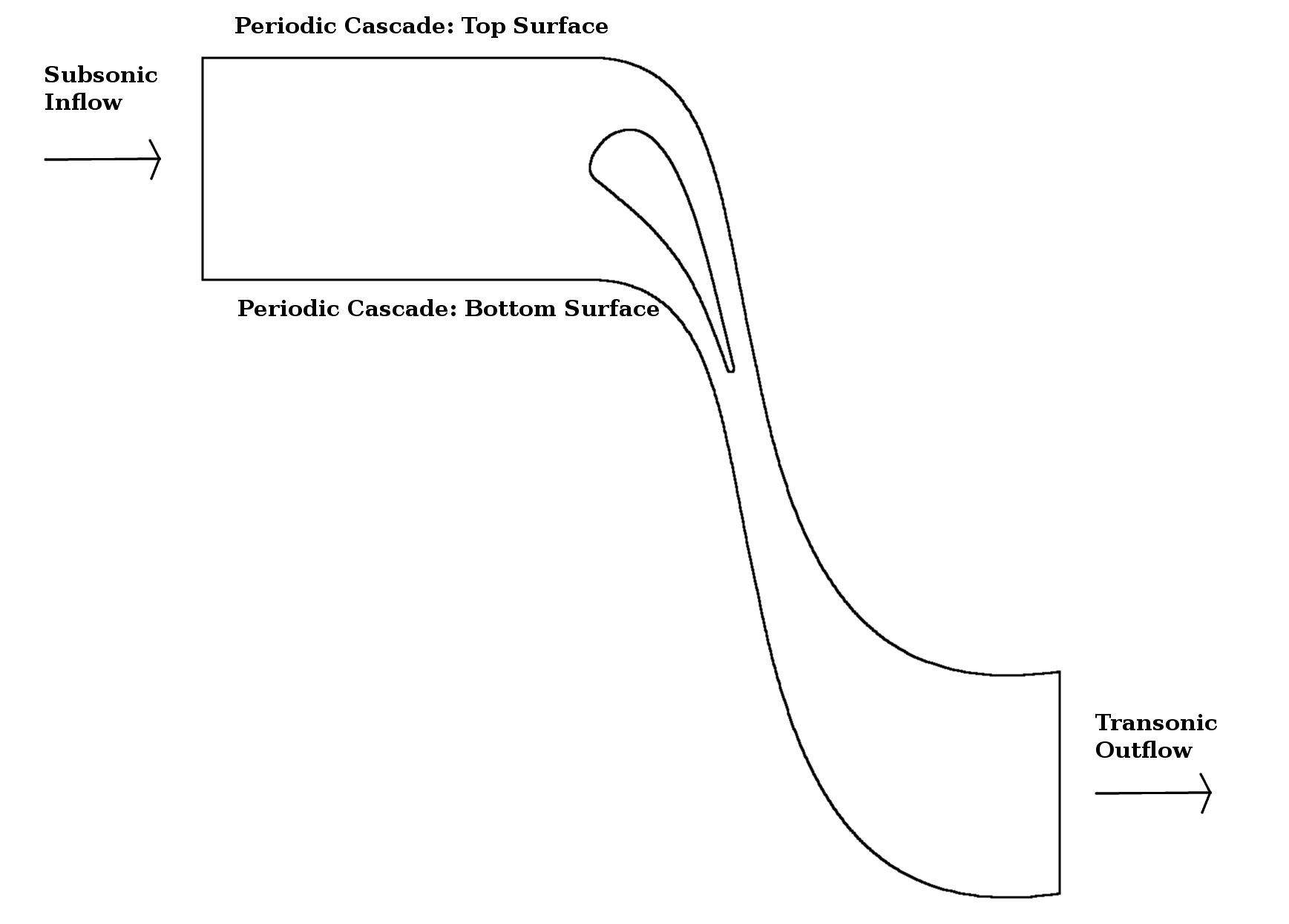}
\end{center}
\caption{Turbine vane computational domain}
\label{f:vane_full_geom} 
\end{figure}

\begin{figure}[htb!]
\begin{center}
    \includegraphics[width=0.65\textwidth,trim={0 0.9cm 0 0},clip]{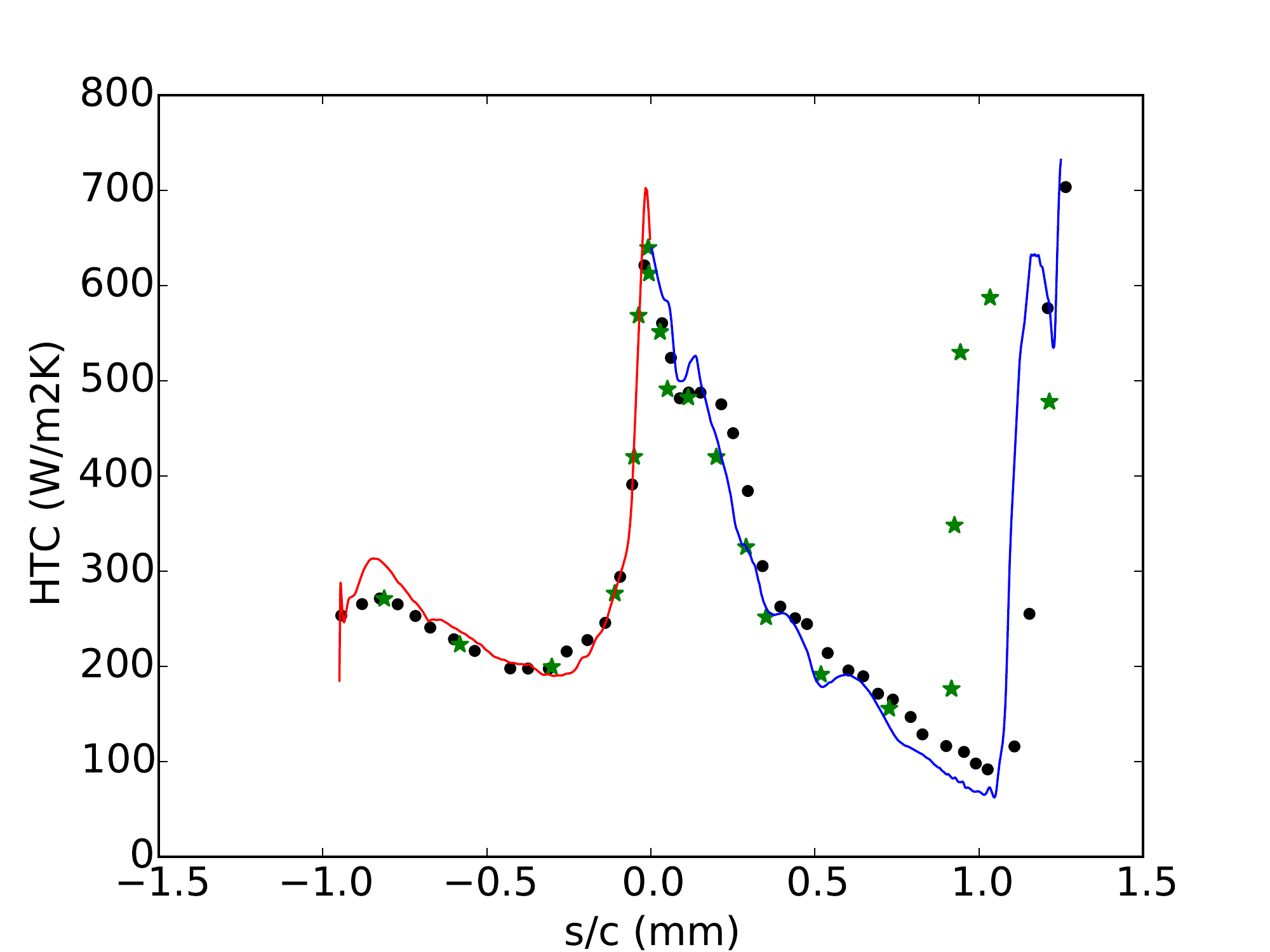}
\end{center}
\caption{Comparison of heat transfer coefficient on the surface of the vane obtained from experimental
    data produced by Arts \cite{arts1992aero} (denoted by black dots),
    numerical data from an LES on a resolved wall grid produced by Gourdain \cite{gourdain2012comparison} (denoted by green colored stars) 
    and
    numerical data generated by an LES on an under-resolved wall grid (denoted by blue and red colored lines)
    at isentropic Mach number 0.9 and Reynolds number $10^6$.
    In the figure, the color blue denotes suction side, the color red denotes pressure side.
    The $x$-axis represents the distance from the leading edge along the surface of the vane normalized 
    by the chord length.}
\label{f:vane_validation1}
\end{figure}

\begin{figure}[htb!]
\begin{center}
    \includegraphics[width=0.65\textwidth,trim={0 0.9cm 0 0},clip]{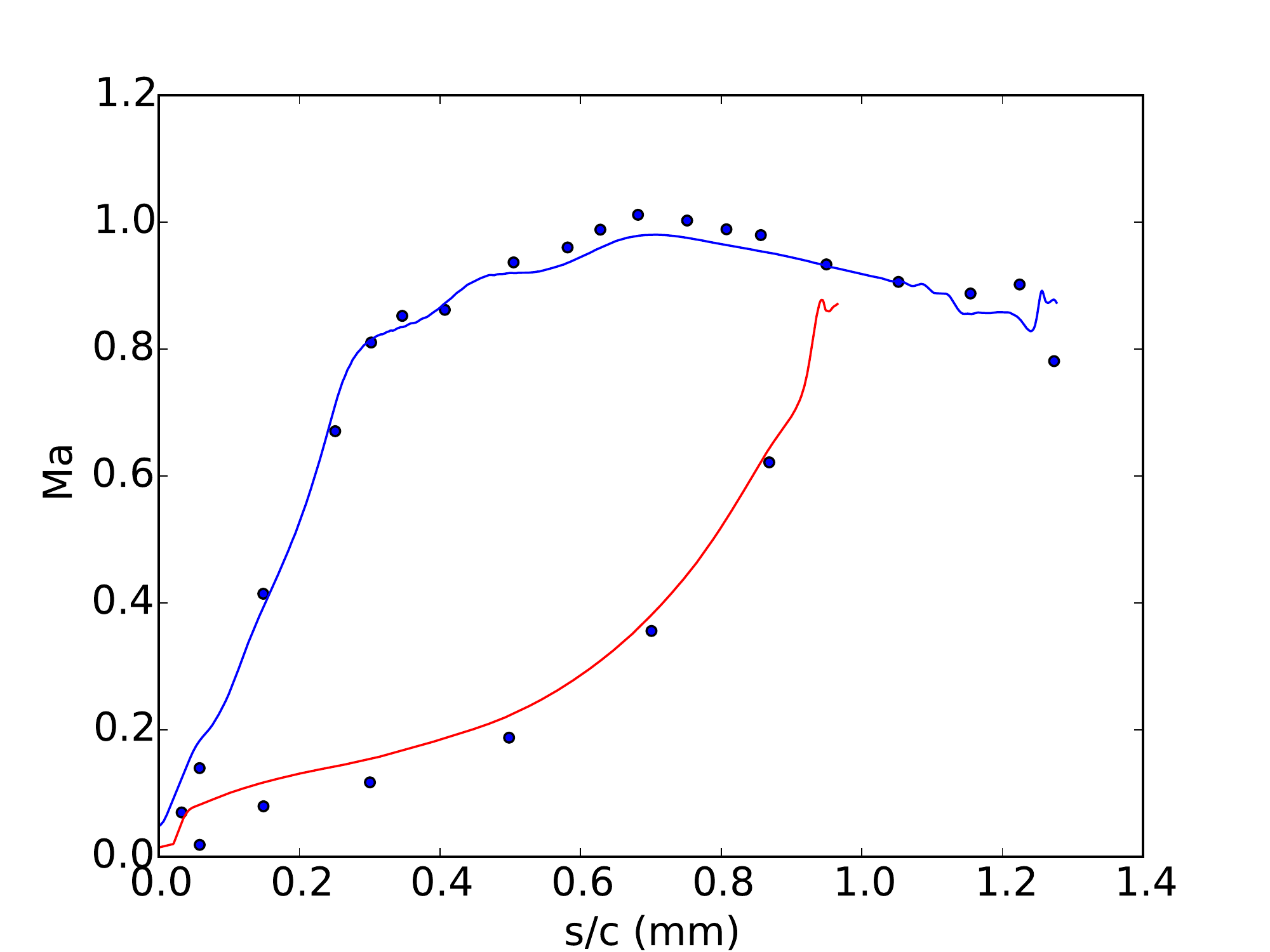}
\end{center}
\caption{Comparison of isentropic Mach number on the surface of the vane obtained from
experimental data produced by Arts \cite{arts1992aero} (denoted by blue dots) and numerical data generated by an
    LES on an under-resolved wall grid (denoted by blue and red colored lines)
    at isentropic Mach number
         0.875 and Reynolds number $10^6$. 
    In the figure, the color blue denotes suction side, the color red denotes pressure side.
    The $x$-axis represents the distance from the leading edge along the surface of the vane normalized 
    by the chord length.
    }
\label{f:vane_validation2}
\end{figure}

\begin{figure}[htb!]
\begin{center}
    \includegraphics[width=0.75\textwidth]{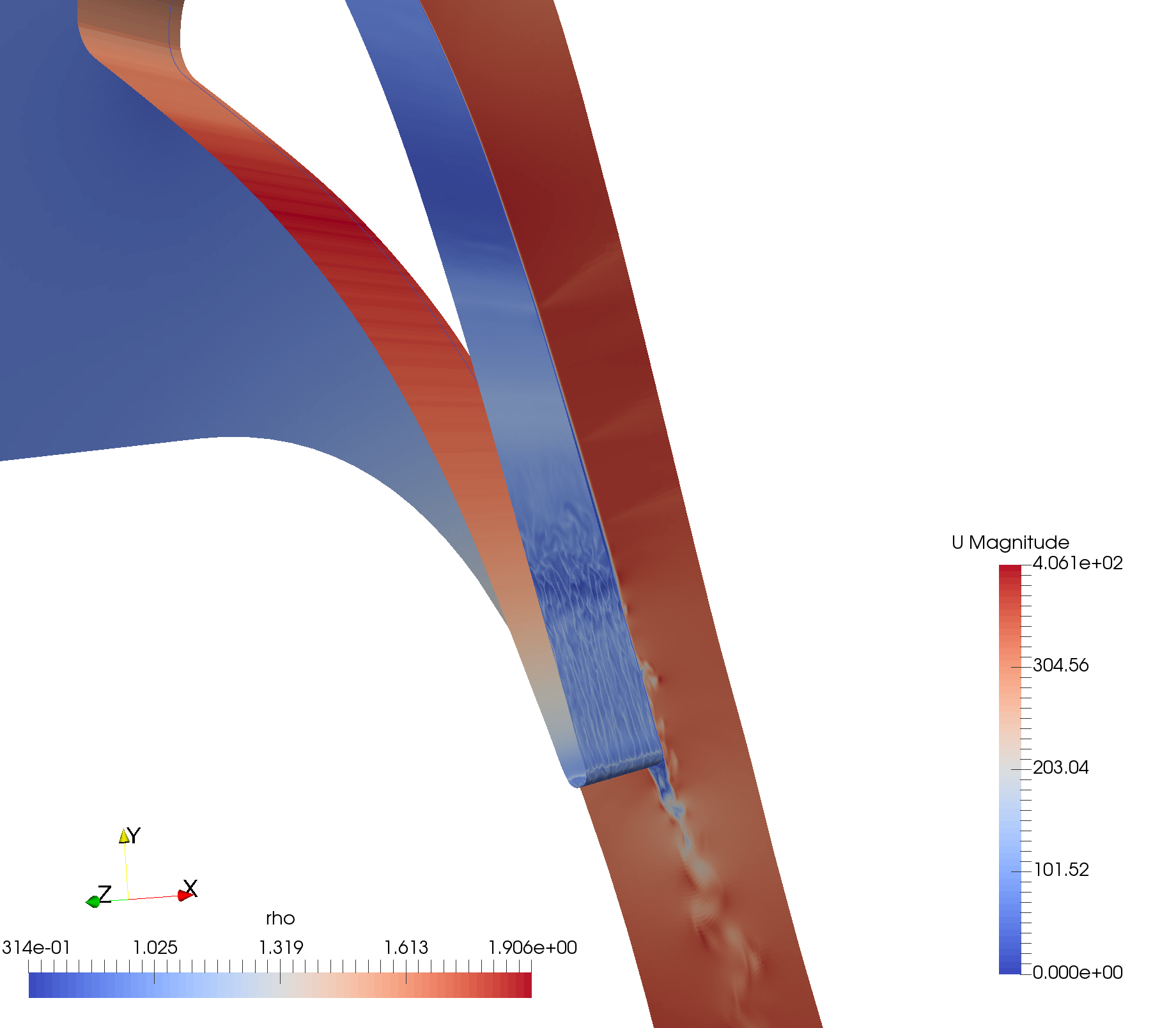}
\end{center}
\caption{An instantaneous snapshot of the flow field near the trailing edge of the
    vane. The density (units: $kg/m^3$) is colored 
    over the surface of the vane and the magnitude of the velocity field (units: $m/s$) is colored
on the planar cross section.}
\label{f:vane_snapshot} 
\end{figure}

A 2-dimensional slice of the computational domain used to model the flow 
problem is shown in Figure \ref{f:vane_full_geom}.
The $x$-direction is the direction of the inflow, the $y$-direction
is the direction of the periodic cascade from the bottom to the top surface and
the $z$-direction is perpendicular to the $x$ and $y$-directions. 
The tip of the leading edge serves as the origin of the computational domain.
Periodic boundary conditions are used in directions transverse to the flow direction.
Static pressure equal to $1\,atm$ is prescribed on the outlet boundary.
A non-reflecting characteristic boundary condition is used
on the inlet boundary of the domain \cite{poinsot1992boundary}.
The stagnation pressure (computed from the downstream isentropic Mach number) and
stagnation temperature ($420\,K$) are prescribed on the inlet boundary. Using this
information on one side of the boundary and the solution fields from the interior of
the domain on the other side, the
Roe approximate Riemann solver is used to propagate the fluxes on the inlet boundary.
The surface of the vane is maintained at a constant temperature of $300\,K$.
The span-wise extent is approximately $0.16\,c_l$
, which is sufficient to accurately capture 
most of the important flow features \cite{gourdain2012comparison}. 
In order to resolve the small scale eddies
of the flow near the wall, the dimensionless wall normal cell spacing ($y_{+}$)
needs to be below $1$ wall unit \cite{choi2012grid}. 
However, this paper does not adopt this practice due to computational
cost considerations.
As the Reynolds number of the flow is high, the $y_{+}$
restriction along with a maximum CFL number of $1.2$ over the 
domain of the flow problem,
significantly lowers the maximum time step size that can be used by an explicit 
time integration scheme. 
To obtain results from the flow simulation in a reasonable time frame,
the maximum $y^+$ is kept at $10$. 
This does not appear to 
lead to significant loss in physical accuracy of the simulation as
demonstrated by the comparison of the experimental
heat transfer and static pressure data, obtained by Arts \cite{arts1992aero}, with numerical data
generated by an LES on an under resolved wall grid
in Figures \ref{f:vane_validation1} and \ref{f:vane_validation2}. 
The LES on the under resolved wall grid appears to provide better
results than the LES on a resolved wall grid. This potentially
can be attributed to the difference in sub-grid scale models,
the former uses an implicit LES model while the latter uses 
the Wall-Adapting Local Eddy Viscosity (WALE) model which could be introducing
modelling error in the solution.
The maximum $x^{+}$ is $250$ and 
maximum $z^+$ is $50$. The total number of cells in the mesh of the 
computational domain is approximately $16$ million.
A visualization of the magnitude of an instantaneous velocity field of the flow 
over the turbine vane near the trailing edge is shown in Figure \ref{f:vane_snapshot}.


\subsection{Design objective}
\label{s:optim_objective}
\begin{figure}[htb!]
\begin{center}
    \includegraphics[width=0.8\textwidth]{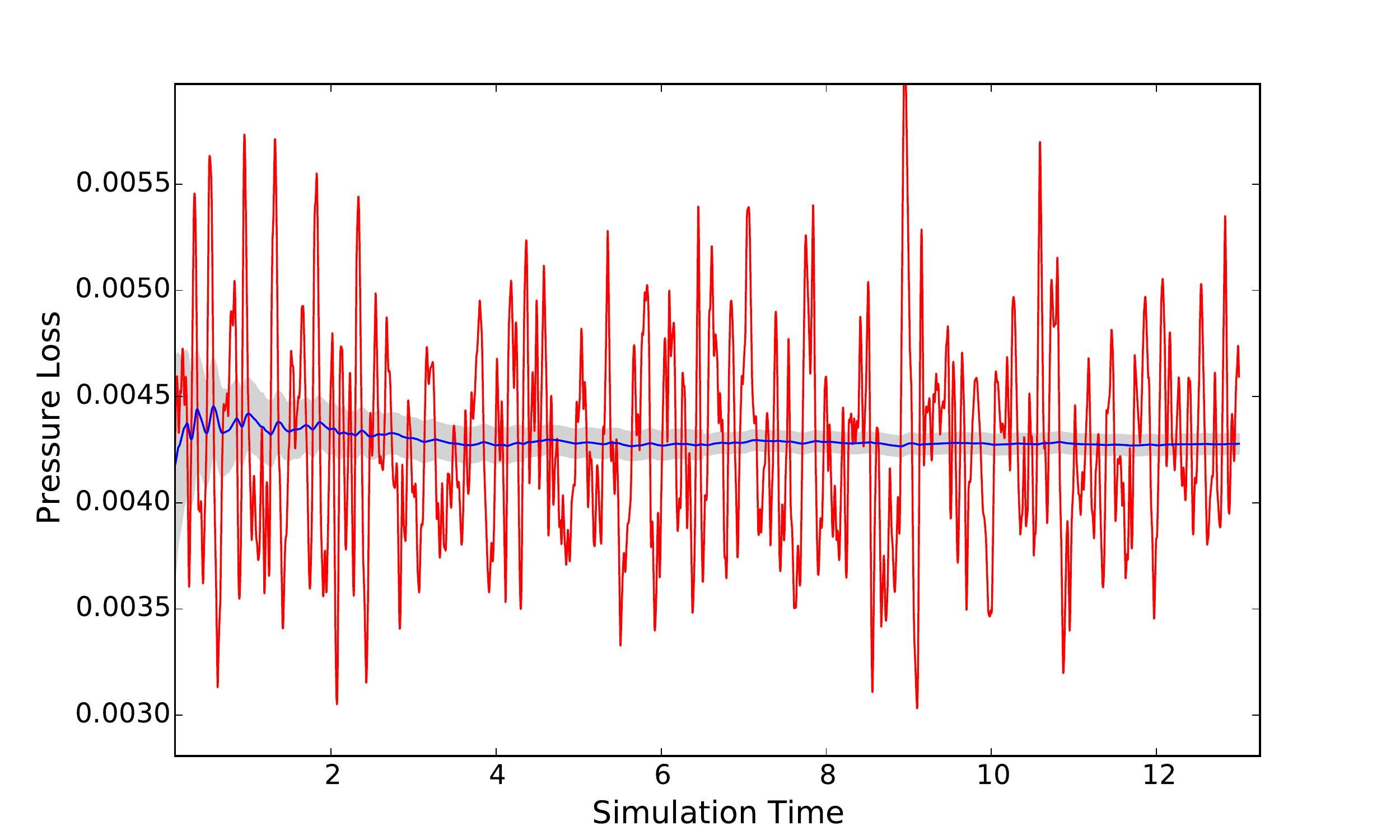}
\end{center}
    \caption{Scaled instantaneous pressure loss coefficient ($b(\bar{p}_l)$) plotted as a function of time (represented 
    by time units).
The blue time series denotes the cumulative mean and the gray shaded area denotes a single standard
    deviation of the sample mean. The procedure for time averaging is discussed in Section \ref{s:optim_objective}.}
\label{f:vane_pt_history} 
\end{figure}

The design objective for the trailing edge shape optimization is 
a linear combination of the stagnation pressure loss downstream
of the vane and heat transfer near the trailing edge of the vane.   
The stagnation pressure loss is represented by the infinite time-averaged
and mass flow-averaged stagnation pressure loss coefficient ($\bar{p_l}$)
\begin{align}
\begin{split}
    \label{eq1}
    \bar{p_l} &= \frac{\bar{p}_{t,l}}{p_{t,in}} \\
    \bar{p}_{t,l} &= \lim_{t_e\to\infty}\frac{1}{t_e}\int_{0}^{t_e} 
    \frac{\int_{S_p} \rho_pu_n(p_{t,in}-p_{t,p})\,dS_p}{\int_{S_p} \rho_p u_n\,dS_p}\, dt \\
    p_{t,p} &= p_p (1 + \frac{\gamma-1}{2}M_p^2)^{\frac{\gamma}{\gamma-1}} \\
    p_{t,in} &= p_{ex} (1 + \frac{\gamma-1}{2}M_{is}^2)^{\frac{\gamma}{\gamma-1}} 
\end{split} 
\end{align}
where $M_p$ is the Mach number on the plane, 
$p_p, \rho_p, p_{t,p}$ are the pressure, density and stagnation pressure 
on the plane respectively, $S_p$ represents
the plane surface, $u_n$ is the velocity normal to the plane surface,
$p_{t,in}$ is the inlet stagnation pressure,
$p_{ex}$ is the exit static pressure and
$M_{is}$ is the downstream isentropic Mach number.
A visualization
of the time history of the instantaneous pressure loss is shown in Figure \ref{f:vane_pt_history}.
\begin{figure}[htb!]
\begin{center}
    \includegraphics[width=0.8\textwidth]{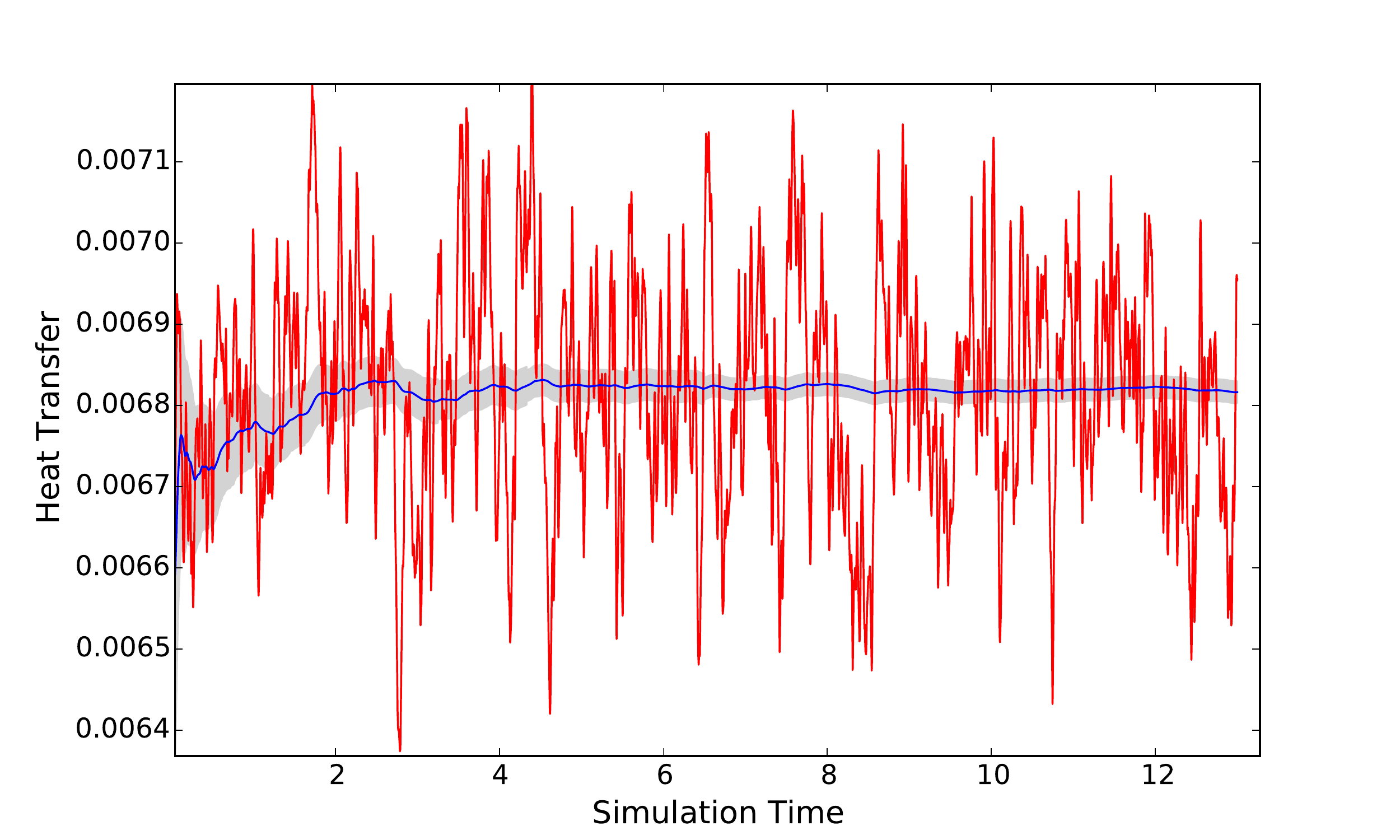}
\end{center}
    \caption{Scaled instantaneous Nusselt number ($a(Nu)$) plotted as a function of time (represented by time units).
The blue time series denotes the cumulative mean and the gray shaded area denotes a single standard
deviation of the sample mean. The procedure for time averaging is discussed in Section \ref{s:optim_objective}.}
\label{f:vane_htc_history} 
\end{figure}
The heat transfer is represented by the Nusselt number ($Nu$).
\begin{equation}
    Nu = \frac{\bar{h}L}{k}
\end{equation}
where $k$ is the thermal conductivity at $T = 300\,K$, $k = 0.028 \,\frac{W}{mK}$, $L$ is the
trailing edge radius for the baseline case, $0.0105\,c_l$, and $\bar{h}$ is the time-averaged
heat transfer coefficient over a part of the vane starting from $0.414 \, c_l$ 
downstream of the leading edge in the direction of the inflow and leading up to the tip of the trailing edge. 
The equation for $\bar{h}$ is 
\begin{equation}
    \bar{h} = \lim_{t_e\to\infty}\frac{1}{S_vt_e\Delta T}\int_{0}^{t_e} \int_{S_v} k \frac{\partial T}{\partial n}\, dS_v\, dt,
    \label{eq2}
\end{equation}
where $S_v$ is the surface area and $\Delta T = 120\,K$ is the temperature difference 
between the surface of the blade and the stagnation temperature of the flow.
A visualization
of the time history of the instantaneous heat transfer is shown in Figure \ref{f:vane_htc_history}.
The design objective for the optimization is a linear combination of two quantities,
the Nusselt number and pressure loss coefficient, and is given by
\begin{equation}
    \bar{J} = a(Nu) + b(\bar{p}_l)
\end{equation}
where $a = 5\times 10^{-4}$ and $b = 0.4$. The  values for $a$ and $b$
are chosen such that the contribution of both $Nu$ and $\bar{p}_l$ to
the sum is equal for the baseline case.

As the fluid flow solution is obtained for a finite time,
the infinite time-averages in Equations \ref{eq1} and \ref{eq2}
are approximated using a finite time-average. 
The length of the time averaging interval
is chosen to provide a statistically converged estimate of the
infinite time average. A converged estimate is an estimate that
has a standard deviation (or standard error) that is less than $10\%$ of the magnitude 
of the estimate itself.
Through a numerical investigation it is found to be equal to 
$N=6$ time units.
$1$ time 
unit ($t_r$) is the time the flow takes to travel from the inlet
boundary to the outlet boundary.
\begin{equation}
    t_r = \frac{1.5c_l}{u_r}
\end{equation}
This 
interval is sufficient to obtain the design objective with
approximately $5\%$ standard error relative to the estimate.

A procedure is required to compute an estimate for the finite time-average
from an instantaneous time history of the design objective.
The finite time-average
can be modeled as a random variable.
The mean estimate of this random variable can be computed using a sample average, 
$\tilde{J}_N$. 
\begin{equation}
    \tilde{J}_N = \frac{1}{N} \sum_{n=N_0}^{N} J_n
\end{equation}
where $J_n$ represents the instantaneous design objective at time step $n$.
The time averaging
is started after an initial transition period to allow for
any transient effects in the time history to 
settle down into a statistical steady state. 
The transition period is determined by finding the time it takes for a running 
time-average to lie within $1$ standard deviation of the full interval time-average.
For the design objective of the trailing edge shape optimization problem, this period
is approximately $N_0=1$ time units. It is large enough
to encompass the transition period for different trailing edge shapes.

The standard deviation of the mean estimate (or sample mean) provides an indication of the 
amount of error in the finite time-average approximation of the design objective.
The standard statistical formula for computing the standard error
of a sample mean cannot be used as the 
instantaneous values of the design objective
are not independent and form a correlated time series. 
Oliver (2014) \cite{oliver2014estimating} suggested using
data fitted autoregressive models to get the correlation function for the time
series. This approach is adopted in this paper. 
Autoregressive models can be written in the following form
\begin{equation}
    J_n = \sum_{i=1}^{p}a_iJ_{n-i} + \epsilon_n
\end{equation}
where $a_i$ are the constant coefficients of the autoregressive model,
$p$ is the order of the model and $\epsilon_n$ are independent, identically
distributed normal random variables,
$\epsilon_n \sim N(0, \sigma^2)$.
The coefficients of the model are determined using the Burg estimation algorithm
with the $CIC$ criterion to decide the model order \cite{burg1982estimation}.
Even though the actual model for the time series may not belong to the 
class of linear stochastic models, the variance computed using
the model's correlation function can be utilized to provide a reasonable estimate
of the variance of the sample mean.
\begin{equation}
    Var(\tilde{J}_N) \approx \frac{Var(J_n)}{N_{d}}
\end{equation}
\begin{equation}
    N_d = \frac{N}{1+2\sum_{k=1}^{\infty} \rho(k)} 
\end{equation}
where $\rho(k)$ is the autocorrelation between $k$ time steps and $N_d$
is the effective sample size.

\subsection{Design parameterization}
Parameterizing the trailing edge of the turbine vane
is a challenging task. For the shapes to be valid, they
have to satisfy a convexity condition about the chord line.
Such a condition is required in order to ensure that 
there are no undulations or obstructions in
the surface of the vane which can increase flow instability
or cause back flow. 
The convexity condition imposes a nonlinear constraint on the shape parameters,
increasing the complexity of the design optimization problem.

\begin{figure}[htb!]
\begin{center}
    \includegraphics[width=0.3\textwidth]{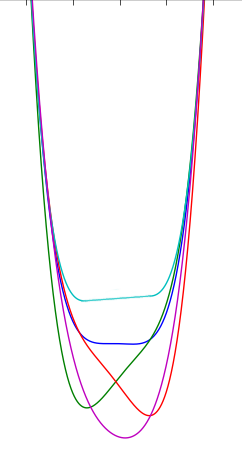}
\end{center}
\caption{Visualization of the 5 basis shapes}
\label{f:vane_shapes}
\end{figure}

One way to get around the nonlinear constraint is to 
parameterize the trailing edge of a 2-dimensional turbine vane 
as a linear combination of $5$ trailing edge shapes, with the 
weights serving as the parameters.
If the candidate trailing edge shapes are convex, then they form a reduced basis of
all convex shapes \cite{samareh2001survey}, ensuring
that the parameterized trailing edge shape is convex.
The linear combination is implemented as a weighted average of the 
2-dimensional coordinates of
the basis shapes. 
If the coordinates of the basis shape $j$ are denoted by $x^j_i$,
then the coordinates of a new shape are given by
\begin{equation}
    x_i(\bm{\alpha}) = \sum_{j=1}^{5} \alpha_j x^j_i
\end{equation}
The sum of the weights ($\alpha_j$) is equal to 1. 
Hence, the weights $\alpha_j$ of the $5$ basis shapes satisfy
the following constraint,
\begin{equation}
    \sum_{j=1}^{5} \alpha_j = 1
\end{equation}
The above equality constraint can be transformed into
an inequality constraint by eliminating one of the $\alpha_j$ variables.
This results in a formation of a $4$-dimensional parameterization
of the trailing edge shape with the following
inequality constraint,
\begin{equation}
    \sum_{j=1}^4 \alpha_j \le 1
\end{equation}
The $5$ basis shapes are chosen
such that the parameterized shapes explore a large subset of convex shapes. 
They are shown in Figure \ref{f:vane_shapes}.
One of the basis shapes is from the baseline turbine vane design from VKI.
The $5$ basis shapes are linearly independent, meaning that there are
no two sets of parameters $\bm{\alpha}$ that generate the same shape. 

Evaluating the design objective and gradient for a trailing 
edge shape requires the generation of a new mesh for obtaining the flow and adjoint
solutions. The mesh corresponding to the parameterized shape is formed
by perturbing the mesh of the baseline design.
The mesh generation process starts by projecting the nodes of the
mesh of the baseline design onto the parameterized design by
minimizing distance between the nodes of the respective designs.
Once projected, a displacement vector is computed from the nodes
of the baseline mesh nodes on the surface of the vane
to parameterized mesh nodes.
This displacement vector on the vane surface is propagated through
the remaining mesh using a linear elasticity model for the mesh nodes \cite{sokolowski1992introduction,dwight2009robust}.
This model can be used under the assumption that the magnitude
of the displacement vector is small. The linear elasticity equations are 
\begin{align}
    \nabla \cdot \bm{\sigma} &= 0 \\
    \bm{\sigma} &= \lambda \epsilon_{ii}\bm{I} + 2\mu\bm{\epsilon} \\
    \bm{\epsilon} &= \frac{1}{2}(\nabla \bm{u} + \nabla \bm{u}^T)
\end{align}
where $\bm{u}$ is the displacement, $\lambda$ and $\mu$ are
constants which govern the elasticity of the mesh nodes for propagating
the displacement. The linear elasticity equations are solved 
using the boundary condition $\bm{u}=\bm{c}$ for
the mesh nodes on the surface of the vane, where $\bm{c}$ is the displacement vector
of the parameterized design
from the baseline design.

The gradient for the new trailing edge shape is evaluated using the
viscosity stabilized adjoint method, which is described in Section
\ref{s:adjoint}.
The function that maps the parameters of the shape to
the location of the parameterized mesh nodes is difficult to differentiate.
Hence, instead of directly providing the gradient of the design objective with respect
to the shape parameters, the adjoint method provides the gradient 
with respect to discrete mesh fields like cell centers, mesh
normals, et cetera, which are typically used to represent
the mesh in finite
volume methods. These gradients
are denoted by $\pd{J}{m_i}$. The finite difference method is 
used to obtain the gradient with respect
to the parameters of the shape ($\pd{J}{\alpha_j}$).
\begin{equation}
    \pd{J}{\alpha_j} \approx \pd{J}{m_i}\frac{(m_i(\bm{\alpha} + \bm{\epsilon_j}{})-m_i(\bm{\alpha}))}{\epsilon}
\end{equation}
where $\bm{\epsilon_j}$ is a zero vector with
the same dimension as $\bm{\alpha}$ and with the $j^{th}$ entry
set to $\epsilon = 10^{-5}$.
Hence, for each design objective gradient evaluation, a set of $4$ meshes
are generated by perturbing each shape parameter by $\epsilon$ separately. 
The corresponding perturbations in the discrete mesh fields 
for each of the shape parameters are
utilized to obtain $\frac{(m_i(\bm{\alpha} + \bm{\epsilon_j}{})-m_i(\bm{\alpha}))}{\epsilon}$.
Multiplying this quantity with $\pd{J}{m_i}$ obtained from
the adjoint method and summing over all the indices $i$ provides the design objective gradient.

\section{Adjoint-based design optimization}
\label{s:adjoint}
The adjoint method for physics-based numerical simulations
is a widely used tool for accelerating the engineering
design optimization process 
by efficiently obtaining the gradient
of the design objective function with respect to the
design parameters. The adjoint method
requires a single additional solution field (known as the adjoint solution) for computing
the gradients, in comparison to the finite difference method
of computing gradients which requires $n$ additional solution fields
for $n$ design parameters \cite{giles1997adjoint}.

\subsection{Viscosity-stabilized adjoint method}
Application of the adjoint method to LES 
produces large magnitude adjoint solutions 
which eliminates it's effectiveness. 
This is especially true
when the design objective is a long-time averaged
quantity \cite{blonigan2012towards}.
The divergence of the adjoint solution for turbulent 
fluid flows is explained by the chaotic dynamics of
turbulence \cite{lapeyre2002characterization,maurer1980effect}.
Like many chaotic systems, turbulent flows
exhibit the butterfly effect \cite{strogatz2014nonlinear}, which means that the
instantaneous flow solution fields are sensitive
to perturbations in the design parameters. The exponentially growing difference in the solution
fields causes the adjoint method to incorrectly compute the gradients for long-time
averaged quantities.
The authors have proposed a solution to this problem
by adding minimal amounts of localized artificial 
viscosity to the adjoint equations \cite{talnikar2016,talnikarphd} that
is sufficient to stabilize the adjoint solution.
Viscosity is added only in regions where there is a high likelihood of  divergence of the adjoint flow field.
Numerical results have shown that even with the artificial viscosity that
stabilizes the adjoint solution, the 
accuracy of the gradients is maintained within 10-20\% of the true
design objective gradient. 

\begin{figure}[htb!]
    \centering
        \includegraphics[width=0.7\linewidth]{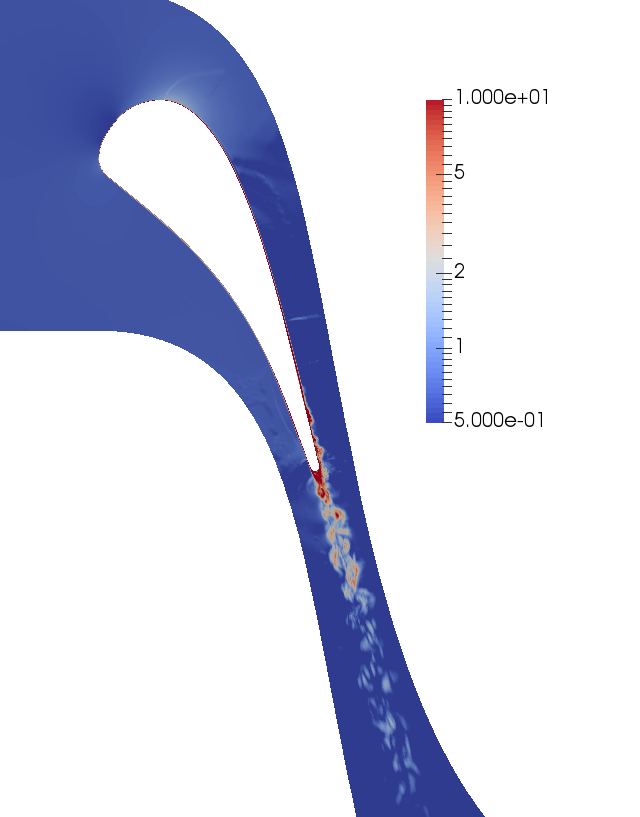}
    \caption{Visualization of the scalar field, $\lambda_1$}
    \label{f:symmetrizer_field2}
\end{figure}

The regions of divergence of the adjoint solution are found by
performing an weighted $L_2$ norm or energy analysis of the Turkel symmetrized adjoint equations for 
compressible fluid flows \cite{talnikarphd}. An indicator field that shows
the regions of divergence is given by the maximum generalized eigenvalue ($\lambda_1$)
of $\frac{1}{2}(\mathbf{M} + \mathbf{M}^T)$ and $\mathbf{N}$ where $\mathbf{M}$
and $\mathbf{N}$ are matrices defined at every point
of the domain of the flow problem and are
given by
\begin{align}
    \begin{split}
    \mathbf{M} &= \mathbf{M_1-M_2}\\
        \mathbf{M_1} &= \frac{1}{2}\begin{pmatrix}
        \divU & \nabla_1 c & \nabla_2 c & \nabla_3 c & 0 \\
        \nabla_1 c & \divU & 0 & 0 & 0 \\
        \nabla_2 c & 0 & \divU & 0 & 0 \\
        \nabla_3 c & 0 & 0 & \divU & 0 \\
        0 & 0  & 0 & 0 & T_r^2\divU \\
    \end{pmatrix} \\
    \mathbf{M_2} &= \begin{pmatrix}
        \frac{\gamma-1}{2}\divU & \frac{1}{\rho c}\nabla_1 p & \frac{1}{\rho c}\nabla_2 p & \frac{1}{\rho c}\nabla_3 p & \frac{T_r^2}{2\rho c}\divU \\
        \frac{\gamma-1}{2\rho c}\nabla_1 p & \nabla_1 u_1& \nabla_2 u_1& \nabla_3 u_1 & \frac{T_r^2}{2\gamma p\rho } \nabla_1 p \\
        \frac{\gamma-1}{2\rho c}\nabla_2 p & \nabla_1 u_2& \nabla_2 u_2& \nabla_3 u_2& \frac{T_r^2}{2\gamma p\rho} \nabla_2 p \\
        \frac{\gamma-1}{2\rho c}\nabla_3 p & \nabla_1 u_3& \nabla_2 u_3& \nabla_3 u_3& \frac{T_r^2}{2\gamma p\rho} \nabla_3 p \\
        0 &  (\nabla_1 p - c^2 \nabla_1 \rho)  &  (\nabla_2 p - c^2 \nabla_2 \rho)  &  (\nabla_3 p - c^2 \nabla_3 \rho) & 0
    \end{pmatrix} \\
    \mathbf{N} &= 
    \begin{pmatrix}
        1 & 0 & 0 & 0 & 0 \\
        0 & 1 & 0 & 0 & 0 \\
        0 & 0 & 1 & 0 & 0 \\
        0 & 0 & 0 & 1 & 0 \\
        0 & 0 & 0 & 0 & T_r^2
    \end{pmatrix}
    \end{split}
\label{e:turkel}
\end{align}
where $T_r$ is a dimensional constant quantity given by
\begin{equation}
    T_r = C_t\frac{p_r}{u_r}
\end{equation}
where $C_t$ is a non-dimensional tunable factor that can be in the range $(0, \infty)$.
For simplicity, $C_t$ is set to $1$.
Figure \ref{f:symmetrizer_field2} shows a visualization of the
scalar field, $\lambda_1$, for flow over a turbine vane. 
The scalar field is normalized
by its spatial $L_2$ norm. 
The scalar field has a large
magnitude in the trailing edge region of the suction side of the turbine vane.
This is to be expected from a physical understanding of the problem
as this is
the region where the turbulent boundary layer is formed. The formation
of a turbulent wake and flow separation leads to the 
divergence of the adjoint solution.

The adjoint solution is damped by adding 
artificial viscosity to the compressible adjoint equations.
Viscosity is only added to the adjoint equations and not the original
flow equations.
The amount of viscosity added is proportional to $\lambda_1$. 
The adjoint equations with the artificial viscosity are
\begin{equation}
        -\frac{\partial\mathbf{\hat{w}}}{\partial t} - (\mathbf{\hat{A}}_{i}-\mathbf{\hat{A}}^v_{i})\frac{\partial \mathbf{\hat{w}}}{\partial x_i}
     = 
     \frac{\partial}{\partial x_i} (\mathbf{\hat{D}}^T_{ji}
      \frac{\partial\mathbf{\hat{w}}}{\partial x_j}) + 
        \underbrace{\eta\frac{\mu_{r}}{\norm{\lambda_1}_\infty\rho_r}
     \frac{\partial}{\partial x_i} (
     \lambda_1
      \frac{\partial\mathbf{\hat{w}}}{\partial x_i})}_{\text{artificial viscosity term}}
\end{equation}
    where $\eta$ is a tunable non-dimensional scaling factor,
    $\mathbf{\hat{w}}$ represents the adjoint solution
    and $\mathbf{\hat{A}}_i, \mathbf{\hat{A}}^v_i, \mathbf{\hat{D}}_{ji}$
    represent linearized operators of the compressible Navier-Stokes equations.

\begin{figure}[htb!]
    \centering
    \includegraphics[width=0.7\linewidth]{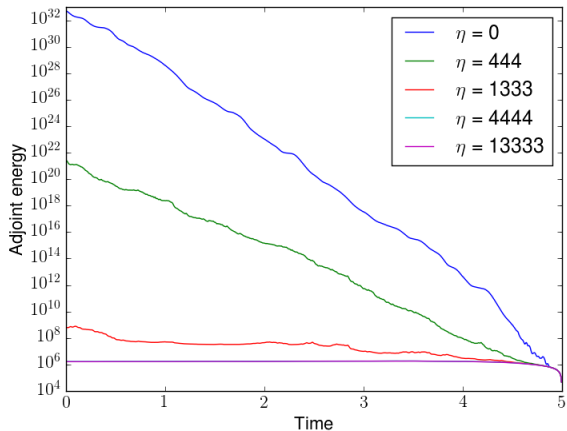}
    \caption{Plot showing the growth of adjoint energy (units: kg s/$\mathrm{m}^3$) for various artificial viscosity scaling factors ($\eta$) as a function
    of time (represented by time units) for the turbine vane problem.}
    \label{f:vane_energy}
\end{figure}

The choice of $\eta$ is important.
Too large values of $\eta$ will result in fast stabilization
of the adjoint solutions, but might adversely affect the accuracy
of the gradients obtained from the adjoint solution. 
Too small values of $\eta$
might not be sufficient to stabilize the adjoint solution.
A reasonable strategy for deciding the appropriate value of $\eta$ 
is to obtain multiple adjoint solutions with different values of
$\eta$ and choose the minimum $\eta$ 
which stabilizes the adjoint
solution over the entirety of the simulation.
After obtaining multiple adjoint solutions with different
values of the scaling factor , the
optimal value of $\eta$ for flow over the
turbine vane is determined to be $\eta \approx 4,400$. 
The adjoint solution is obtained over a finite time interval whose length is
$0.5$ time units.
This value is the minimum value
of $\eta$ for which the adjoint solution is stable
for the baseline design of the trailing edge of the turbine
vane. The value of $\eta$ is kept constant for different parameterized
designs of the trailing edge of the turbine vane. Empirically,
it has been observed that the same value of $\eta$ stabilizes the adjoint solution for
different designs. 
Figure \ref{f:vane_energy} shows the growth of adjoint energy 
for various artificial viscosity scaling factors ranging from $\eta = 0$ to $\eta \approx 13,000$.

\subsection{Optimization method}
Design objectives in LES are often
defined as infinite time averages. But, in practice, as the simulations are performed for
a finite time, there is a sampling error associated with the objective function and gradient
evaluations \cite{oliver2014}. 
This error makes it difficult for the optimization algorithm to know whether a new design
point with a lower objective value is actually better than the current optimal or if the difference
can be attributed to noise. 
Additionally, the computational expense of LES mandates optimizers to 
utilize information from all previous evaluations to decide the next design point
to evaluate and not just the last few design points.
Bayesian optimization is a robust global optimization algorithm that
is suitable for optimizing such noisy and expensive objective functions.

In this class 
of optimization algorithms, a surrogate
model is fit to all the design evaluations using a Gaussian process \cite{rasmussen2006gaussian}. 
The surrogate model explicitly models the amount of noise in the objective
and gradient evaluations. It forms a global approximation of the design
objective utilizing information from all past evaluations.
In Bayesian optimization, the next design point to evaluate is decided
by optimizing a metric which is a scalar function on the design space.
The metric generally depends on the surrogate model and past evaluations. 
The Bayesian optimization loop begins by fitting the surrogate model
to the past evaluations, proceeds to optimize the metric to find the next design point and finally evaluates the design.
The loop terminates when there is no appreciable improvement in
the design objective in consecutive evaluation or the
computational resources to perform the optimization are exhausted.
As the evaluations are noisy, the optimal design 
at the end of the optimization process is the minimum of
the surrogate model and not the design with the minimum objective value in all the evaluations.

Using a stochastic process to model the objective function
helps in quantifying the uncertainty in a surrogate model that is trained 
on noisy evaluations. Additionally, it helps in designing 
a metric for deciding the next evaluation
points that works towards finding the optimal design.
A Gaussian process (GP) is a stochastic process whose mean ($\mu(\bm{x})$) and covariance
functions ($k(\bm{x}, \bm{x}^*)$) are sufficient to completely define it, where
$\bm{x}$ is the point in the design space $\omega$. 
A realization of a GP, evaluated on a set of discrete points $\bm{X}$, is
a sample of a multivariate normal distribution with mean $\mu(\bm{X})$ and covariance
$k(\bm{X}, \bm{X})$. 
If the design objective function is denoted by $f(x)$, 
then the corresponding GP is defined by \cite{rasmussen2006gaussian}
\begin{align}
    \mu(\bm{x}) &= \E[f(\bm{x})], \\
    k(\bm{x}, \bm{x}^{*}) &= \E[(f(\bm{x})-\mu(\bm{x}))(f(\bm{x}^{*})-\mu(\bm{x}^{*}))],
\end{align}
where the true functions $\mu(\bm{x})$ and $k(\bm{x}, \bm{x}^{*})$ for the
design objective are unknown.
An initial estimate for the mean and covariance functions is known as the
prior GP. On observing a few evaluations, the prior GP is updated using
the Bayes rule to form the posterior GP.
A prior mean function that is commonly used is $\mu(\bm{x}) = \mu_f$, where
$\mu_f$ is a constant.
The choice of the prior covariance function restricts the
function space associated with the GP.
As a majority of the design objectives typically observed in CFD are smooth (continuous and infinitely differentiable), 
the squared exponential kernel ($k(\bm{x},\bm{x}^{*}) = \sigma_k^2e^{-(\frac{|\bm{x}-\bm{x}^{*}|}{c_l})^2}$)
is used as the covariance function.  This kernel 
has a few hyperparameters that need to be decided before beginning the optimization process.
The hyperparameter, $c_l$, is the length scale of the kernel that determines how fast the function changes over the design space.
The hyperparameter, $\sigma_k^2$, is the signal variance and it controls how much nearby
points in the design space are correlated.

Assuming additive noise in the design evaluations,
the equation for an objective function and gradient evaluation is written as
\begin{equation}
\bm{y} = \begin{pmatrix} f(\bm{x}) + \epsilon_1(\bm{x}) \\ f^\prime(\bm{x}) + \epsilon_2(\bm{x})\end{pmatrix}
\label{e:first}
\end{equation}
where $\epsilon_1, \epsilon_2$ are random variables representing
the noise in the objective function and gradient evaluations respectively.
$\epsilon_1(\bm{x})$ and $\epsilon_2(\bm{x})$ are
assumed to be independent normal random variables, $\epsilon_1(\bm{x}) = N(0, \sigma_1^2(\bm{x}))$
and $\epsilon_2(\bm{x}) = N(0, \sigma_2^2(\bm{x}))$,
where $\sigma_1^2$ and $\sigma_2^2$ denote the variance and are a function
of the design space.
Independence of $\epsilon_1$ from $\epsilon_2$ can be explained
from the observation that the
artificial viscosity in adjoint equations
decorrelates the time series of the gradient evaluations 
from the time series of the objective evaluations.
Due to a lack of knowledge in the correlation structure
of the gradient components, the noise terms in each of the components
of the gradient evaluation are assumed to be independent.
The noise in each design point evaluation is assumed to be independent of 
other design points as
the correlation between the error due to finite time-averaging of the design objectives for different
points in design space is observed to be zero. This is due to the fact that
the chaotic time series, formed by the instantaneous design objective evaluations,
exponentially diverges from the time series of even neighboring design points.
The variance of the noise term is modeled heteroskedastically,
which means that
the amount of noise in the evaluations is
a function of the design point. The dependence of $\sigma_1^2$ and
$\sigma_2^2$ on $\bm{x}$ is not known beforehand. Hence, they
are modeled as independent logarithmic (log) GPs. This
ensures that the variance term remains positive \cite{kersting2007most}. $\sigma_1(\bm{x})$ is
represented as
\begin{equation}
    log(\frac{\sigma_1^2)}{\mu_{\sigma^2_1}}) \sim GP(0, k_1(\bm{x}, \bm{x}^*))
\label{e:second}
\end{equation}
where $k_1 (\bm{x}, \bm{x}^*) = e^{-\frac{|\bm{x}-\bm{x}^*|^2}{c_l^2}}$, 
$\mu_{\sigma_1^2}$ is the prior mean for the log GP. A similar
expression can be used to represent $\sigma_2(\bm{x})$.

Using the aforementioned noise model, evaluations of the objective function and gradient 
for the design points are used to update
the GP using Bayes rule to form a posterior GP.
Consider a set of sample points $\bm{X}_s$ and the
corresponding function and gradient evaluations $\bm{y}_s$.
The mean ($\bm{\mu}_p$) and covariance ($\bm{\Sigma}_p$) of the posterior process, when evaluated
on a set of evaluation points $\bm{X}$ is given by
\begin{align}
    \begin{split}
    \bm{\mu}_p &= \bm{K}^T(\bm{K}_s + \bm{\Sigma}_s)^{-1}\bm{y}_s, \\
    \bm{\Sigma}_p &= k(\bm{X}, \bm{X}) - \bm{K}^T_*(\bm{K}_s + \bm{\Sigma}_s)^{-1}\bm{K},
    \end{split}
    \label{e:gp_update}
\end{align}
\begin{align}
    \begin{split}
    \bm{K}_s &= \begin{pmatrix} k(\bm{X}_s,\bm{X}_s) & k^\prime(\bm{X}_s,\bm{X}_s)^T \\ k^\prime(\bm{X}_s,\bm{X}_s) & k^{\prime\prime}(\bm{X}_s,\bm{X}_s)) \end{pmatrix},
\bm{K} = \begin{pmatrix} k(\bm{X},\bm{X}_s) \\ k^{\prime}(\bm{X}, \bm{X}_s) \end{pmatrix}, \\
    \bm{\Sigma}_s &= \begin{pmatrix} diag[\sigma_1^2(\bm{X}_s)] & 0 \\ 0 & diag[\sigma_2^2(\bm{X}_s)] \end{pmatrix},
    \end{split}
\end{align}
where $k^\prime$ denotes the derivative of $k$ with respect to the first argument,
$k^{\prime\prime}$ denotes the Hessian of the derivatives with respect to 
the first and second arguments.

The variance estimate of the sample mean
of the design objective and gradient evaluations,
discussed in Section \ref{s:optim_objective}, is used to update the noise GPs
$\sigma_1(x)$ and $\sigma_2(x)$ using an expression similar to (\ref{e:gp_update}) without the gradient
evaluations.

Before starting the Bayesian optimization process, it is important to have a set of
evaluations from which an initial surrogate model can be created \cite{pronzato2012design}.
This step, known as design of experiment (DoE), is also
used to estimate the hyperparameters of the GP. The hyperparameters are estimated
using maximum likelihood estimation \cite{rasmussen2006gaussian}. The
mean quantities, $\mu_f$, $\mu_{\sigma^2_1}$ and $\mu_{\sigma^2_2}$, are estimated using a sample
mean of the design and gradient evaluations and their variance estimates.
The design points for the DoE are chosen from
a random subset of points at the corners of the 4-dimensional hypercube that
contains the design space, midpoints of the edges of the hypercube and the centroid of
the hypercube. The points which do not satisfy the constraints of the design optimization problem are omitted.

In the Bayesian optimization loop, after fitting the surrogate model to 
all the past objective function and gradient evaluations,
a metric function is optimized to find the next design point to evaluate.
The metric can be designed to achieve a certain set of optimization goals
\cite{ginsbourger2008multi,contal2013parallel,swersky2014freeze,picheny2013quantile,hennig2012entropy}. 
For example, a popular optimization strategy involves
exploring the design space in the initial part of the optimization process in order to improve the
quality of the surrogate
and then exploiting the surrogate model by evaluating designs close to the minimum of
the surrogate in order to find the optimal design. 
A metric that reflects this strategy is the expected improvement (EI) criterion \cite{jones1998efficient}. 
It provides a good balance between exploration and exploitation. EI
is defined by the following expression
\begin{align}
    \mathrm{EI}(\bm{x}) = \E[\max(f_{min}-f(\bm{x}), 0)],
\end{align}
where $f_{min}$ is the current estimated minimum of the objective, $f_{min} = \mathrm{min}_{\bm{x}\in\omega} \mu(\bm{x})$.
For GPs, EI has a compact analytical form obtained by integrating over the
expectation \cite{snoek2012practical}
\begin{align}
    \mathrm{EI}(\bm{x}) =
    (f_{min}-\mu(\bm{x}))\Phi\left(\frac{f_{min}-\mu(\bm{x})}{\sigma(\bm{x})}\right) +
    \sigma(\bm{x})\phi\left(\frac{f_{min}-\mu(\bm{x})}{\sigma(\bm{x})}\right),
\end{align}
where $\phi$ is the standard normal density, $\Phi$ is the standard normal distribution function
and $\mu(\bm{x})$ and $\sigma(\bm{x})$ are the mean and standard
deviation of the GP. The point in the design space ($\bm{x}_n$), which maximizes EI,
$\bm{x_n} = \mathrm{argmax}_{\bm{x} \in \omega} \mathrm{EI}(\bm{x})$, 
is chosen as the next
design to evaluate. 

If the evaluations of the objective function are not
noisy, EI can be shown to converge to the global minimum \cite{vazquez2010convergence}
provided the objective function belongs to the class of functions that can be 
represented by the surrogate model. 
But, when the evaluations are noisy, there is no proof of convergence.
Additionally, it has been observed that using the EI metric can cause the optimization process to
get stuck in a local minimum\cite{picheny2010noisy}.  The reason is that the optimizer spends too many evaluations exploiting the GP without 
doing enough exploration.
One possible solution to this problem is to choose $f_{min}$
in such a way that EI is biased towards exploration.
\begin{equation}
    f_{min} = \mathrm{min}_{\bm{x}\in\omega} (\mu(\bm{x}) + \beta\sigma(\bm{x}))
\end{equation}
Increasing $\beta$ leads to a significant increase in $f_{min}$, and consequently EI, for regions
of the design space where $\sigma(\bm{x})$ is high.
Hence, due to the larger EI values in unexplored regions, higher $\beta$
results in more exploration.

\begin{figure}[htb!]
    \centering
    \includegraphics[width=0.75\linewidth]{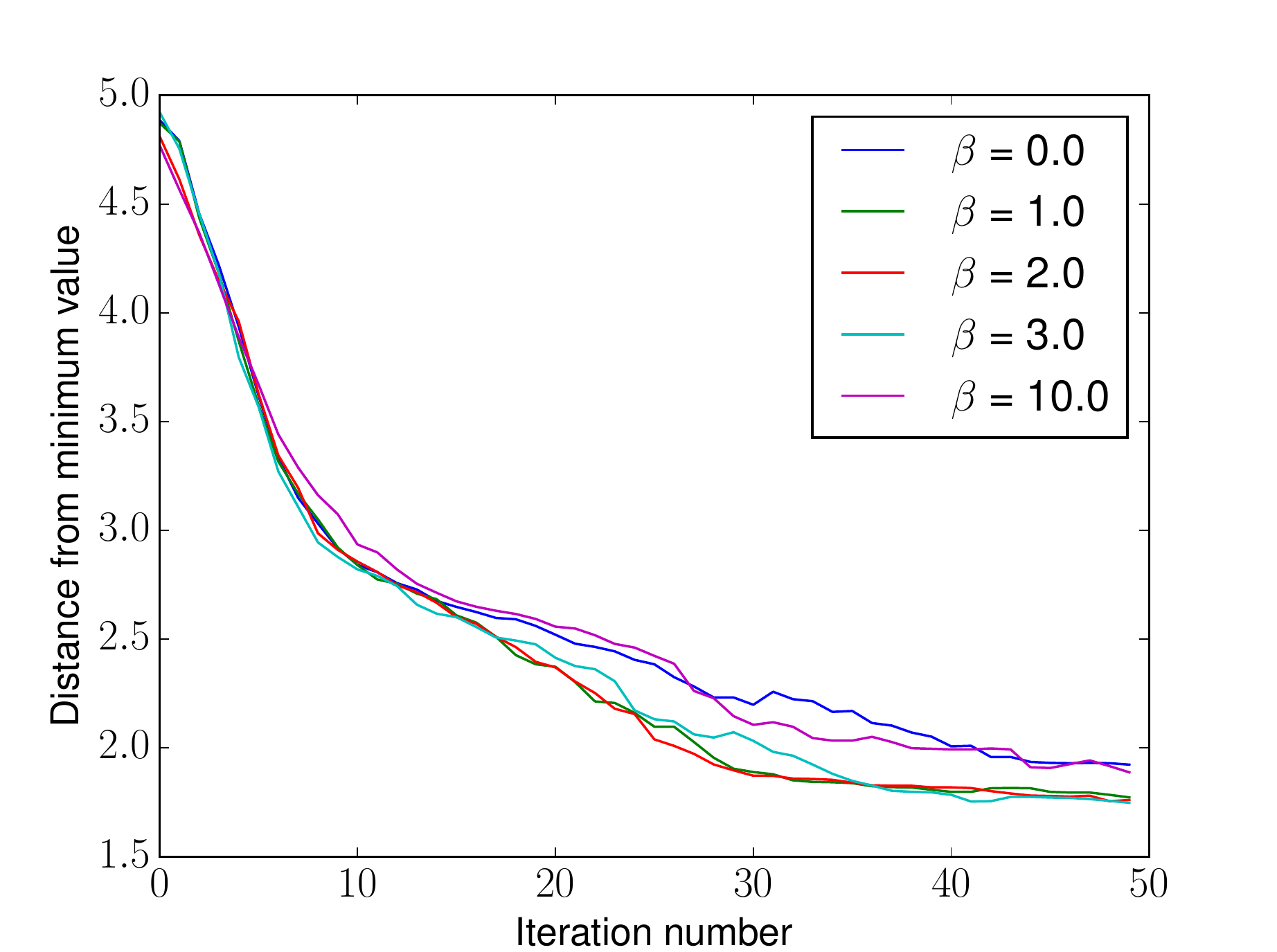}
    \includegraphics[width=0.75\linewidth]{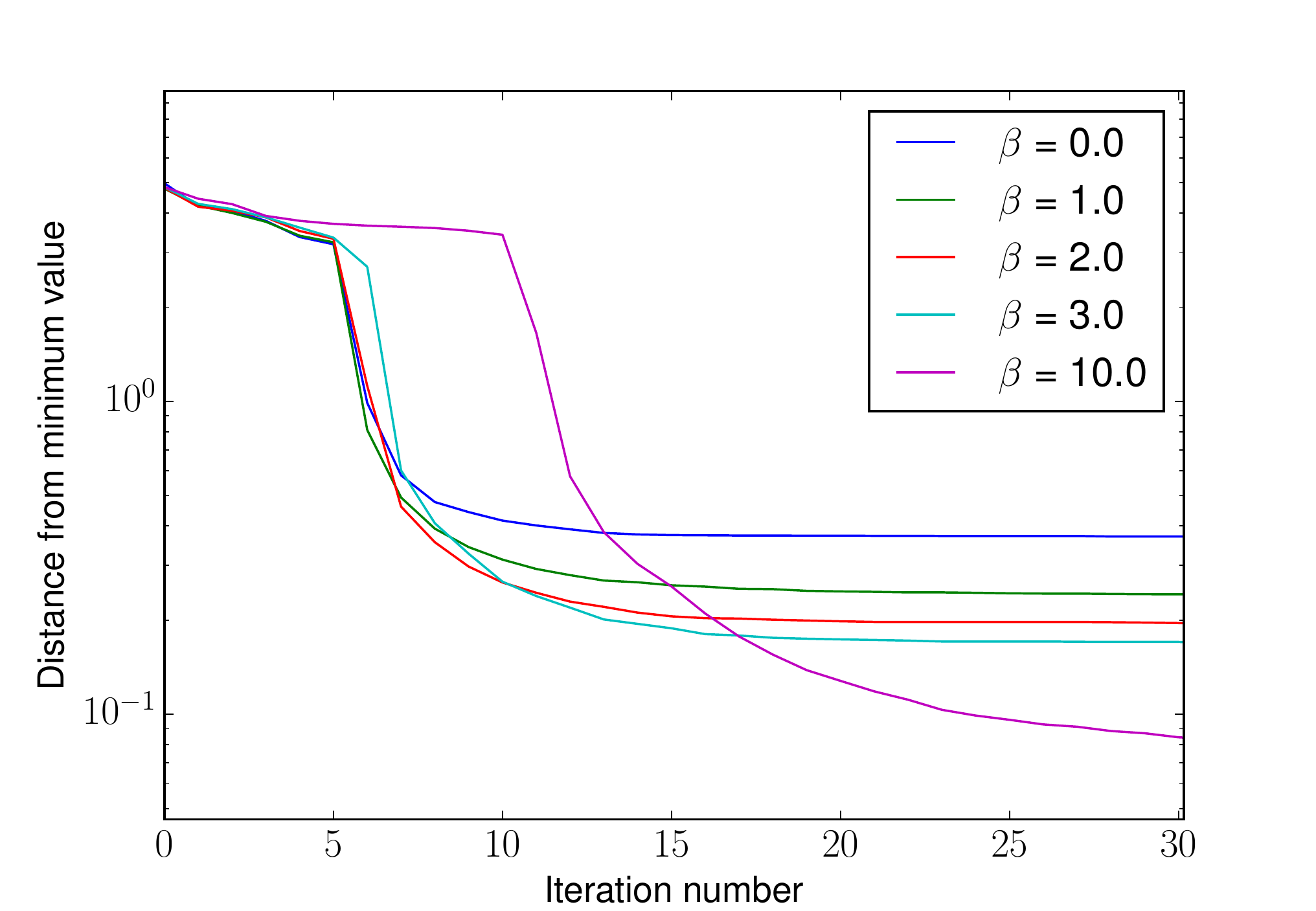}
\caption{Comparison of the distance from the global minimum of the
    minimum design evaluated by the Bayesian optimization algorithm using different 
$\beta$ values for the modified EI metric on the modified noisy Rastrigin function.
    Top figure: Rastrigin function with $\zeta=1.5$.
             Bottom figure: Rastrigin function with $\zeta=0.5$.}
\label{f:ei_beta}
\end{figure}

Numerical experiments show that for various values of $\beta$ the Bayesian optimization
algorithm, with the modified EI metric, converges to the global optimum for a range of noisy functions
including the noisy Rastrigin and long-time averaged quantities of chaotic systems like the Lorenz system.
Figure \ref{f:ei_beta} shows the performance of the exploration biased EI 
metric on 2-dimensional optimization for 
two parameter choices of $\zeta$ for the noisy Rastrigin function.
The function is given by
\begin{equation}
    J(\bm{x}) = -30 + x_1^2 + x_2^2 - 10[cos(2\pi \zeta x_1) + cos(2\pi \zeta x_2)] + \psi z
\end{equation}
where $z$ is the standard normal random variable.
The global minimum of this function is at $\bm{x} = (0,0)$ and the minimum objective value is $-50$.
$\psi$ is set to  $4$, which means that the objective function has a 
large amount of noise.
When $\zeta$ is set to $0.5$,
the objective function has a lower number of local minima than
when $\zeta = 1.5$. 
Figure \ref{f:ei_beta} shows the trajectory of the distance of the minimum evaluation design point
from the global minimum design point, averaged over
$1000$ runs of the optimizer. 
For the objective function with $\zeta=0.5$, 
using any positive value of $\beta$ results in the Bayesian optimization algorithm finding
an optimal design
that is closer to the global minimum than using $\beta = 0$.
Using higher values of 
$\beta$ enables the optimizer to explore more and reach a lower
objective value.
For the objective function with $\zeta = 1.5$, using
$\beta = 1, 2$ or $3$ results in 
better performance for the Bayesian optimization algorithm.
The high number of local minima lowers the convergence rate
of the various optimization algorithms with different
values of $\beta$ and increases the difficulty of finding the global optimal design.
Similar to the $\zeta = 0.5$ case, positive values of $\beta$ causes
the optimizer to suppress exploitation of local minima and do more exploration.
For all the optimization runs, the number of DoE points are set to 4.

\subsection{Algorithm}
The final optimization procedure for the trailing edge shape optimization problem,
where the number of parameters in the optimization problem is $n=4$, is described below
\begin{enumerate}
    \item Evaluate $2n$ design points, where the design points are obtained from
        a design of experiment.
    \item Decide hyperparameters for GP using maximum likelihood estimation applied
        to the $2n$ design evaluations.
    \item Obtain posterior GP by fitting the surrogate model to all the past design evaluations.
    \item Find next design point to evaluate by maximizing the exploration biased EI metric using $\beta=1$.
    \item Evaluate design point.
    \item If the computational resources are exhausted or the optimization process has reached convergence,
        then return the optimal design, else repeat the process starting from step 3.
\end{enumerate}
The cost of this algorithm is $(10n + 5m)C_p$, where $m$ is the number of design evaluations
in the optimization loop and $C_p$ is the cost of obtaining the design objective value (or a single flow solution). 
The cost of a design evaluation is $5C_p$, as the typical cost of obtaining the design objective gradient
(or a single adjoint solution) is $4C_p$.

\section{Results}
\label{s:results}
\begin{figure}[htb!]
\begin{center}
    \includegraphics[width=0.85\textwidth]{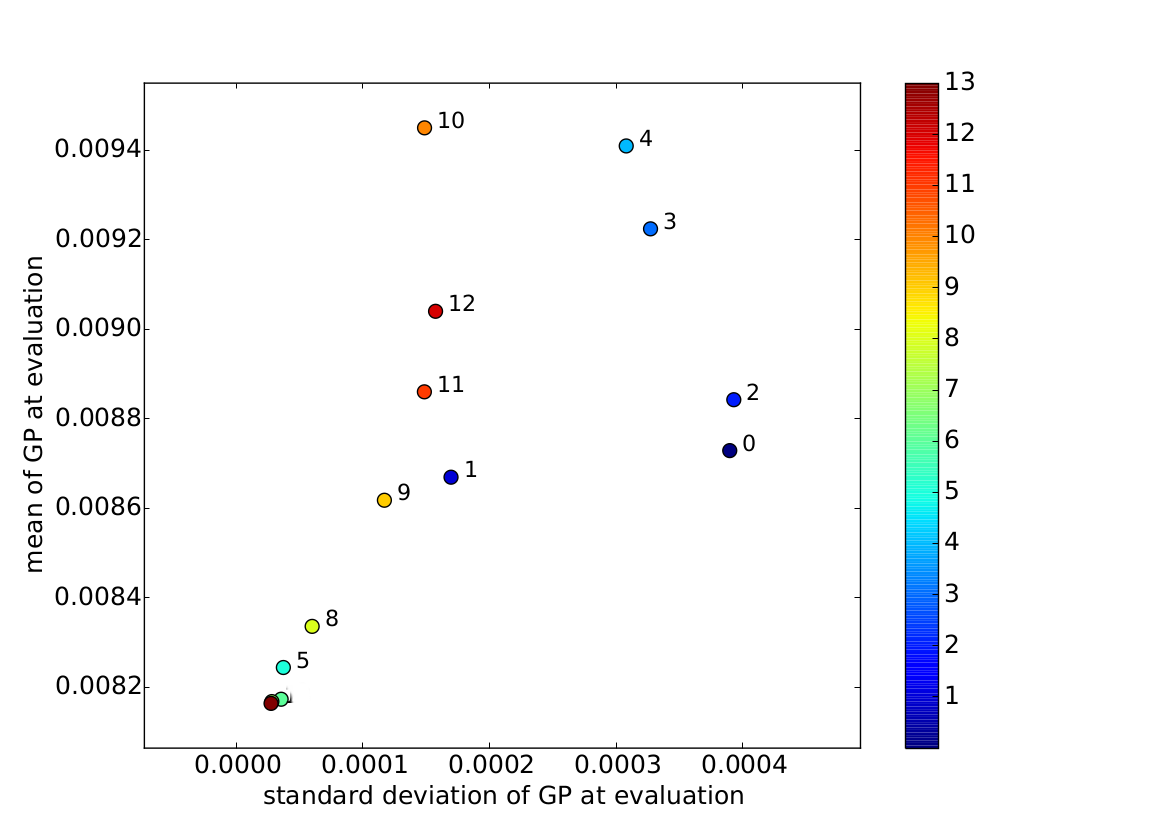}
\end{center}
\caption{Plot of the design objective mean value and standard deviation
    of the posterior GP 
    evaluated at the design point decided by the 
    modified EI metric at each step of the optimization process. The points in the plot are colored 
    and labeled by the optimization step number.}
\label{f:exp_toff}
\end{figure}

The shape of the trailing edge of the turbine vane
is optimized using the modified Bayesian optimization algorithm utilizing
the viscosity stabilized adjoint method for computing gradients.
The optimization process begins with a design of experiment for
$8$ evaluations. 
The hyperparameters for the Gaussian process are 
$c_l = (0.7, 0.4, 0.5, 0.3)$ and $\sigma_k^2 = 10^{-6}$. The prior means 
for the GPs are $\mu_f = 0.0092, \mu_{\sigma^2_1} = 5\times10^{-9}$ and $\mu_{\sigma^2_2} = (10^{-9}, 10^{-7}, 10^{-8}, 10^{-7})$.
The Bayesian optimizer is run
for a total of $16$ design objective and gradient evaluations.
The value of $\beta$ for the exploration biased EI metric is set to $1$.
During the optimization, multiple trailing edge shapes are found
which have a lower design objective value. The optimization ends
when consecutive designs evaluated by the optimizer do not
lead to a reduction in the design objective value or its standard deviation.
Figure \ref{f:exp_toff} shows how the optimizer spends
a majority of the initial evaluations on exploration (high standard
deviation and high mean objective value of the posterior GP at evaluation point) and the evaluations towards
the end of the optimization process in exploitation (low standard
deviation and low to high mean objective value of the posterior GP at evaluation point).

The optimization used a mixture of computational resources consisting of 
CPUs and GPUs. The design objective value was obtained by computing
the numerical flow solution on an Nvidia GeForce GTX 1080Ti, which
has 3584 CUDA cores and 11 GB RAM. The design objective gradient
was obtained by computing the numerical adjoint solution on 
16 Intel Xeon E5-1650 CPUs, where each of the CPUs has 4 cores and 32 GB RAM.
The CPUs are interconnected using Gigabit Ethernet.
The time to solution for a single design objective value is $12$ hours
and for a single design objective gradient is $12$ hours.
Hence, the total computational cost of the optimization
is $18,432$ CPU core hours and $288$ GPU hours.

\subsection{Comparison of optimal and baseline designs}
\begin{figure}[htb!]
\begin{center}
    \includegraphics[width=0.75\textwidth]{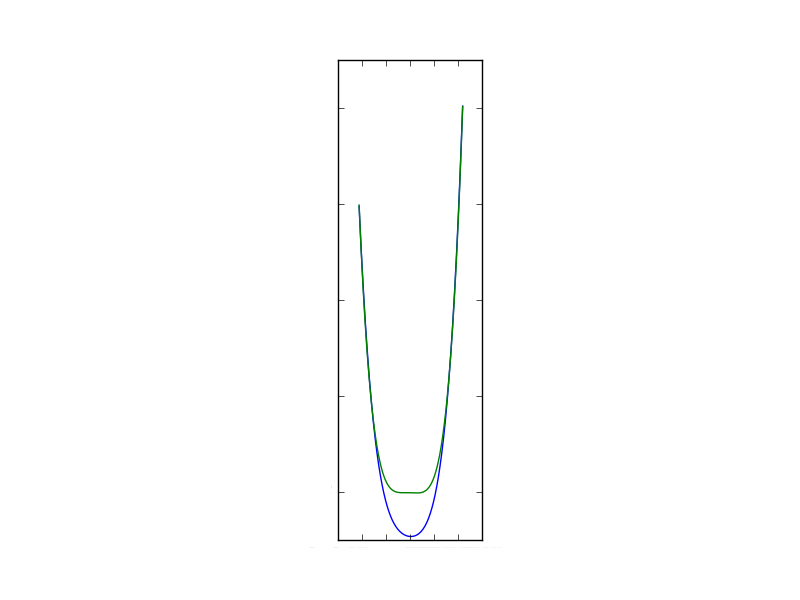}
\end{center}
\caption{Visualization of the current optimal design (blue color) and baseline design (green color)}
\label{f:vane_optimal}
\end{figure}

The optimal design at the end of the optimization process is shown in Figure \ref{f:vane_optimal}.
The baseline design has a design objective value equal to $0.00931 \pm 0.0004$,
whereas the optimal design has the objective value $0.008124 \pm 0.00003$.
The optimal design has an approximately $12\%$ reduction in Nusselt number and $16\%$ reduction in pressure loss coefficient.
The design of experiment procedure produced a design with an objective value $0.00831$. Consequently,
the optimization procedure led to a $2.2\%$  improvement in the design objective over the
design of experiment. 

\begin{figure}[htb!]
    \centering
    \begin{minipage}{.49\textwidth}
        \centering
        \includegraphics[width=1.\linewidth]{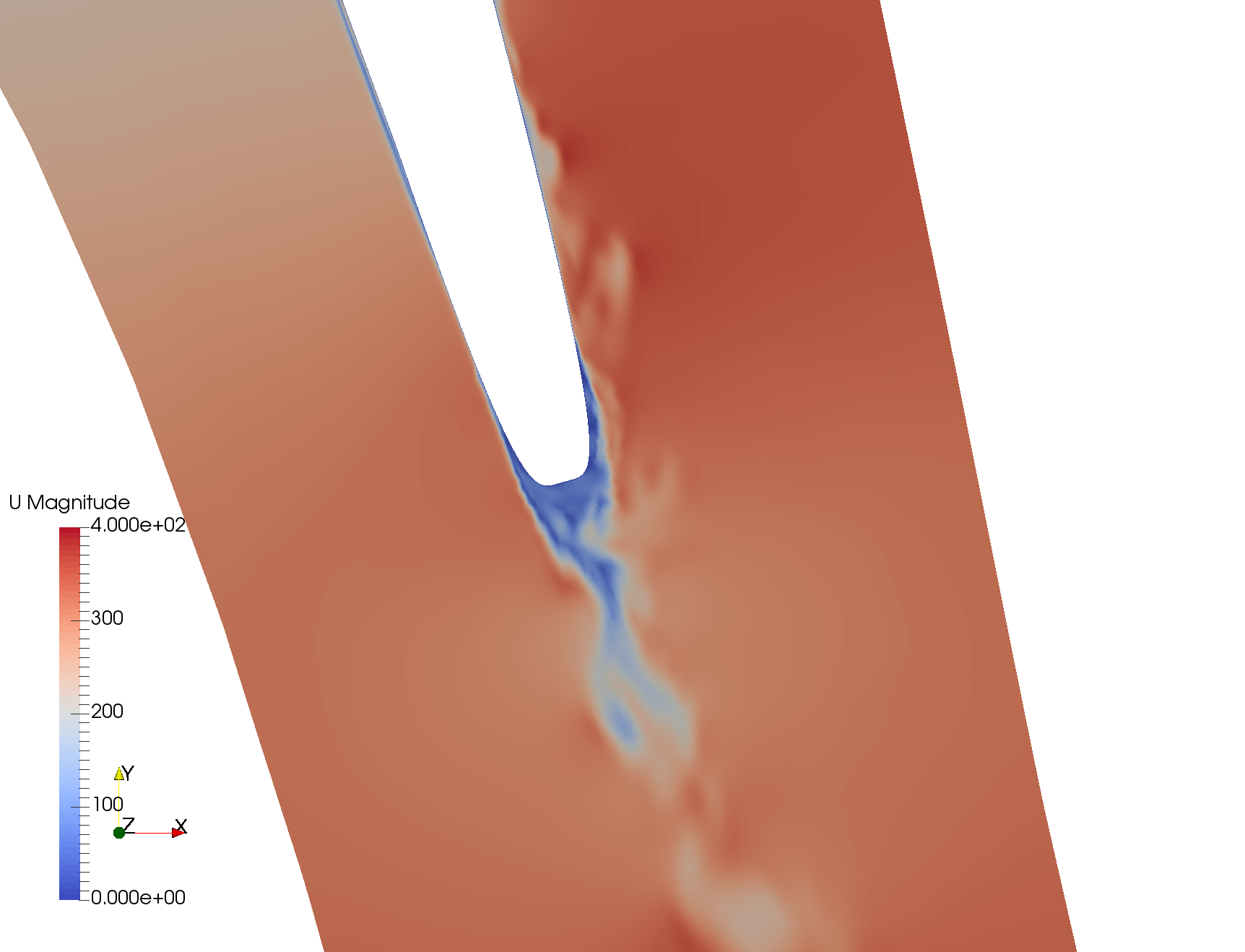}
    \end{minipage}%
    \begin{minipage}{0.49\textwidth}
        \centering
        \includegraphics[width=1.\linewidth]{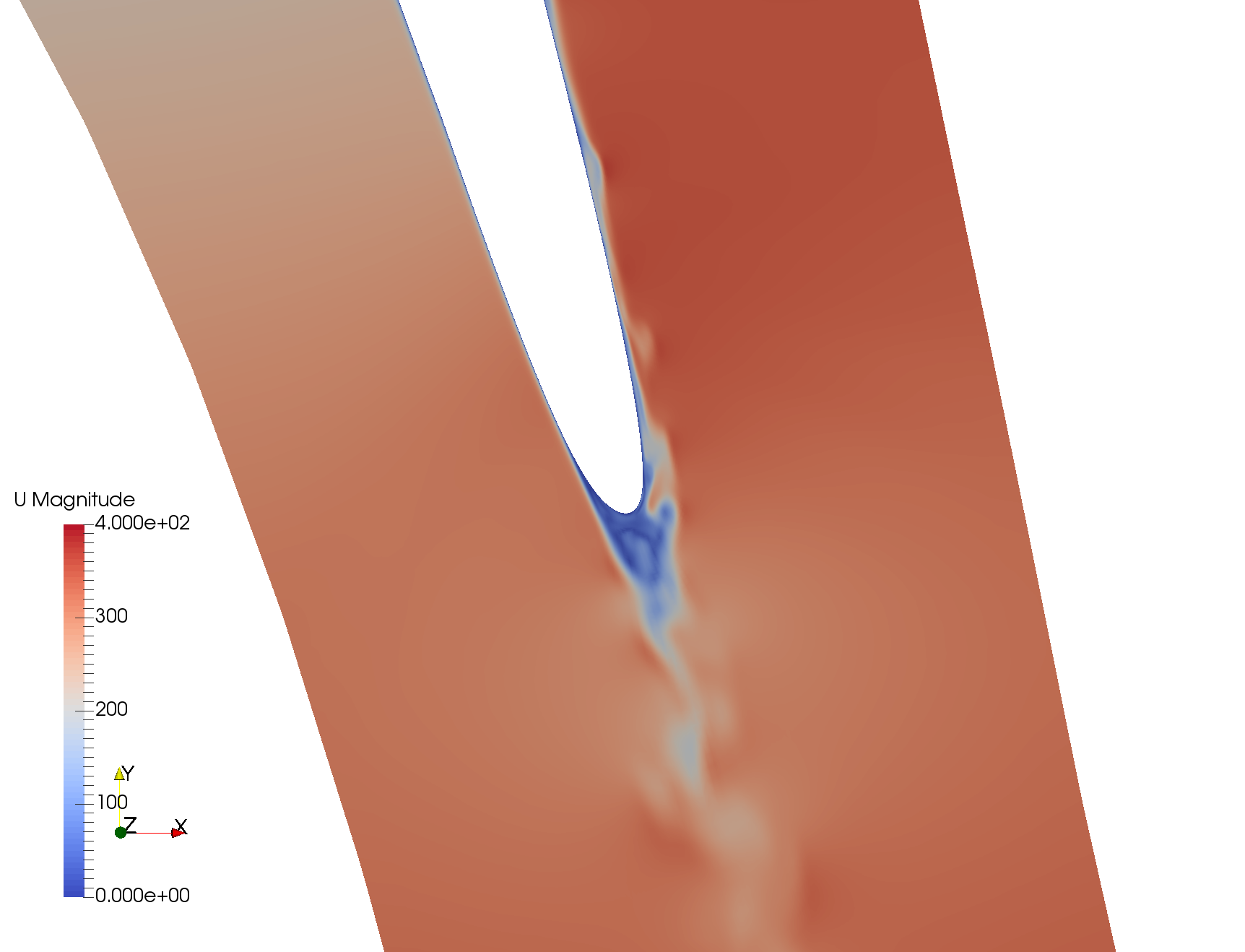}
    \end{minipage}
    \caption{Left figure: Visualization of velocity field for baseline design.
             Right figure: Visualization of velocity field for optimal design}
\label{f:vane_compare_U}
\end{figure}

\begin{figure}[htb!]
    \centering
    \begin{minipage}{.49\textwidth}
        \centering
        \includegraphics[width=1.\linewidth]{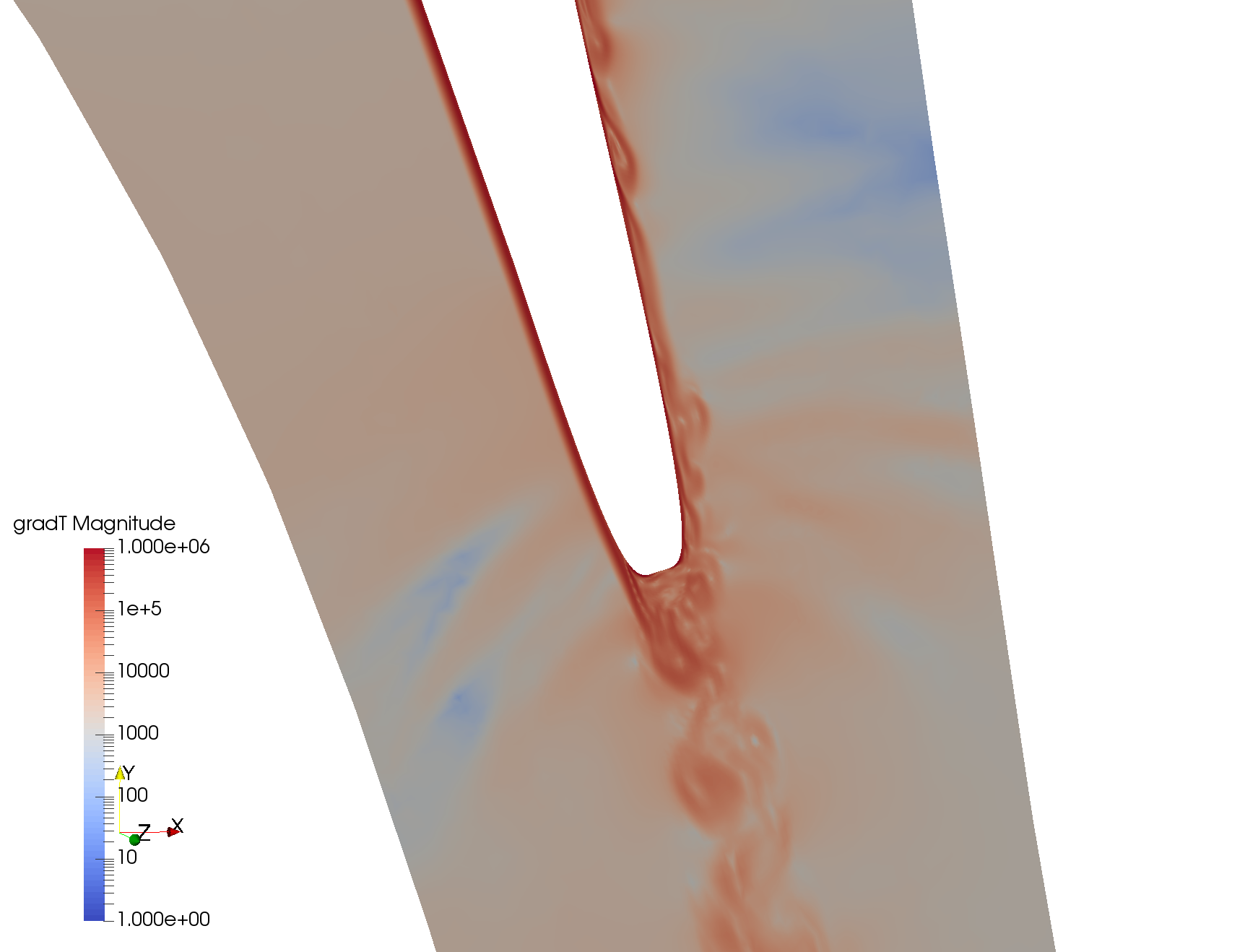}
    \end{minipage}%
    \begin{minipage}{0.49\textwidth}
        \centering
        \includegraphics[width=1.\linewidth]{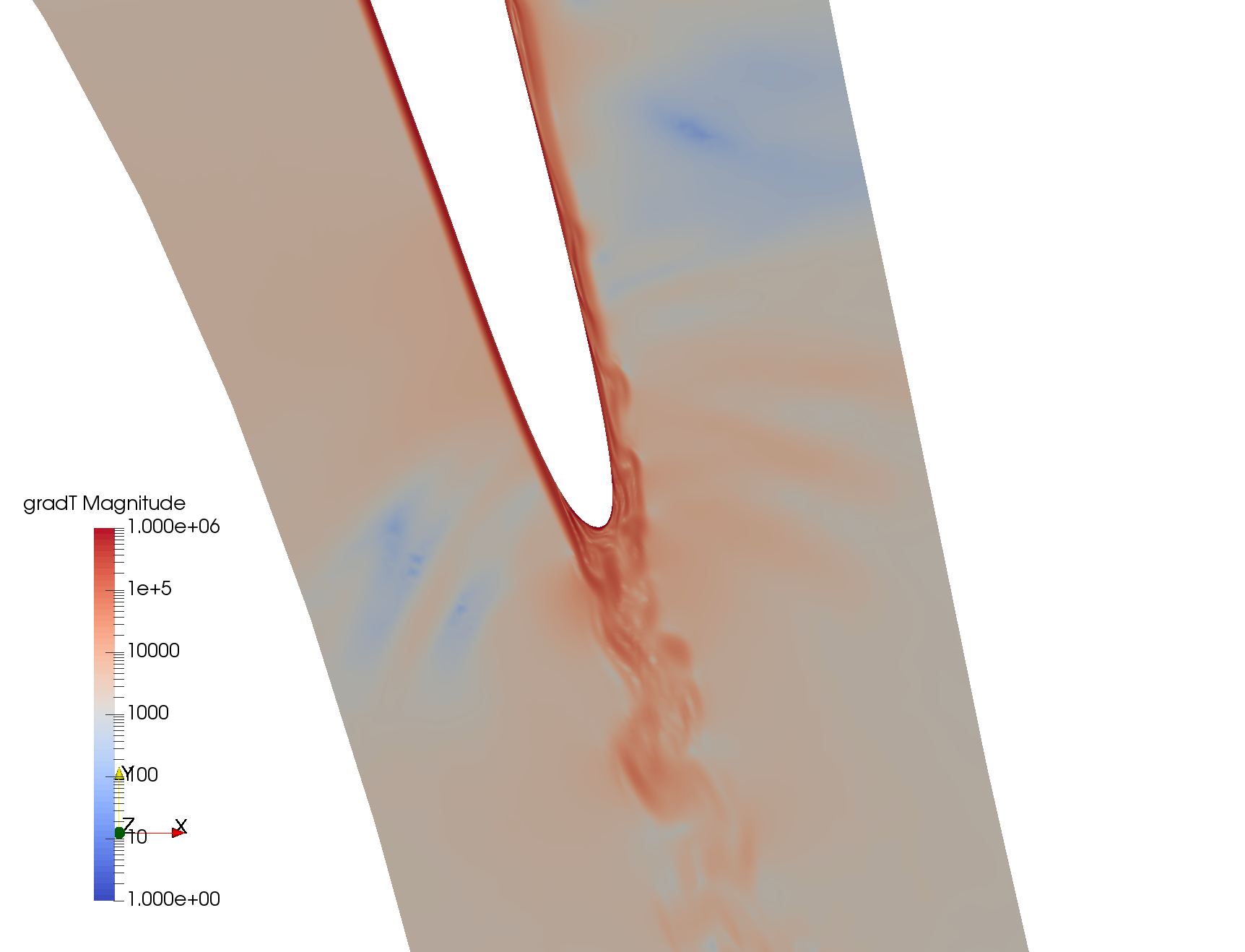}
    \end{minipage}
    \caption{Left figure: Visualization of gradient of temperature field for baseline design.
             Right figure: Visualization of gradient of temperature field for current optimal design}
    \label{f:vane_compare_gradT}
\end{figure}

\begin{figure}[htb!]
        \centering
        \includegraphics[width=0.85\linewidth]{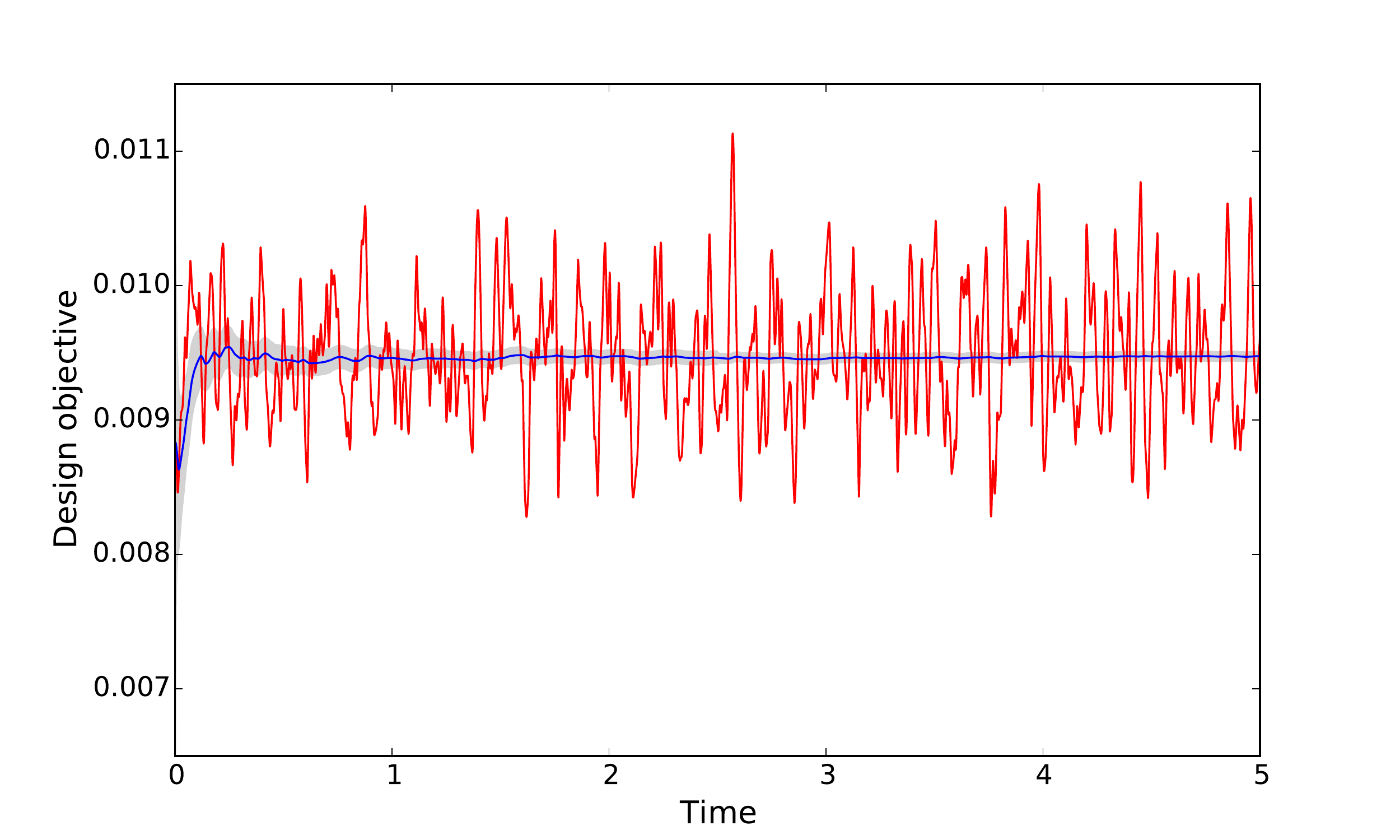}
        \includegraphics[width=0.85\linewidth]{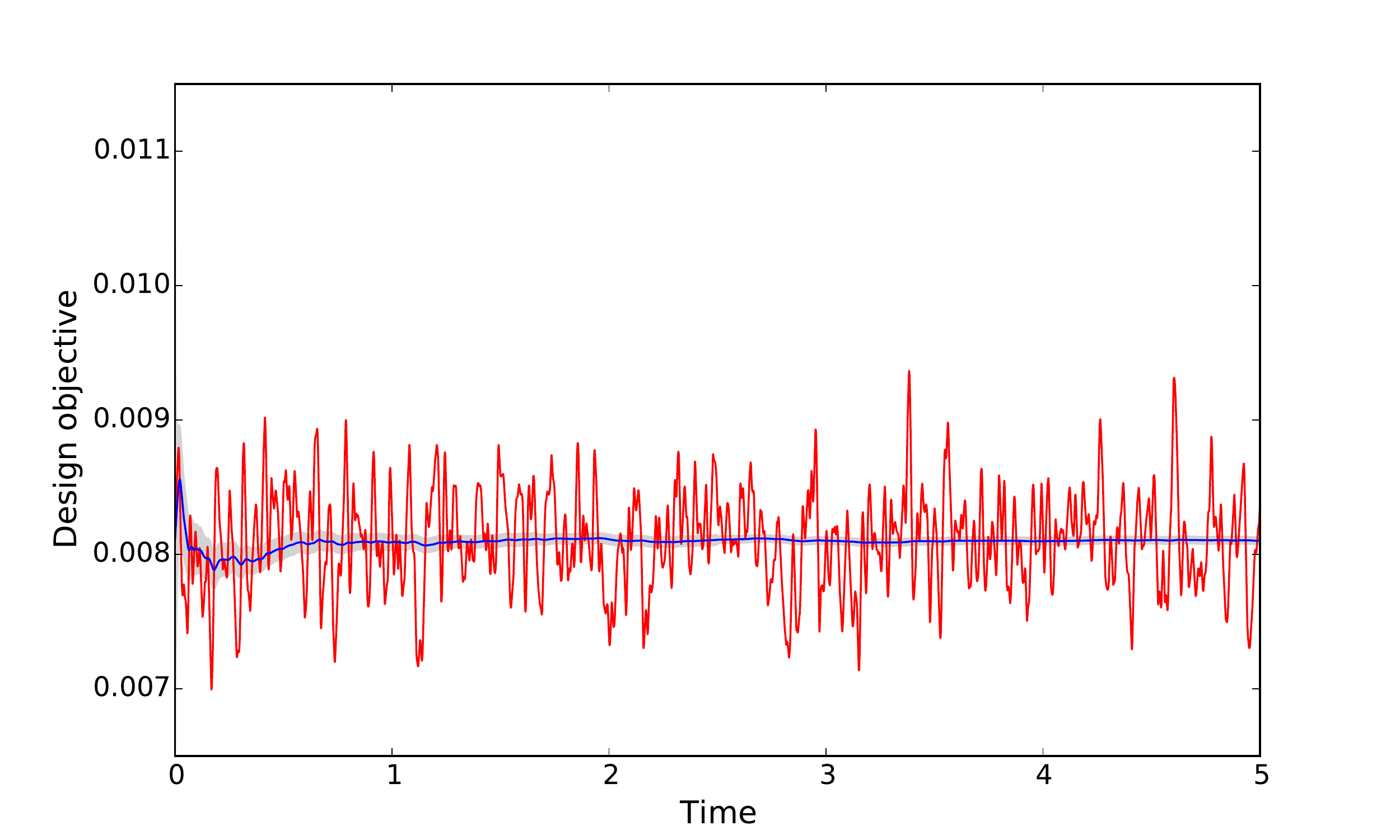}
    \caption{Top figure: Visualization of design objective for baseline design.
             Bottom figure: Visualization of design objective for current optimal design}
    \label{f:vane_compare_objective}
\end{figure}

A visualization of the magnitude of the instantaneous velocity field (U) and the gradient of the temperature field (gradT)
are shown in Figure \ref{f:vane_compare_U} and \ref{f:vane_compare_gradT} respectively.
A comparison of the two fields for the
baseline and optimal design shows that the optimal design 
has a slightly thinner turbulent boundary layer on the suction side
near the trailing edge. The width of the wake 
and the size of the vortex structures in the wake are
smaller in the optimal design.
Both these factors contribute
to the lower stagnation pressure loss in the optimal design. 
A thicker turbulent boundary layer leads to a higher convective 
heat transfer due to more mixing in the flow. Hence, the optimal 
design has a lower heat transfer than the baseline design.
Finally, visualizations of the instantaneous design objective for the 
baseline and optimal designs are shown in Figure
\ref{f:vane_compare_objective}.

\subsection{Comparison to Bayesian optimization without gradients}
\begin{figure}[htb!]
\begin{center}
    \includegraphics[width=0.85\textwidth]{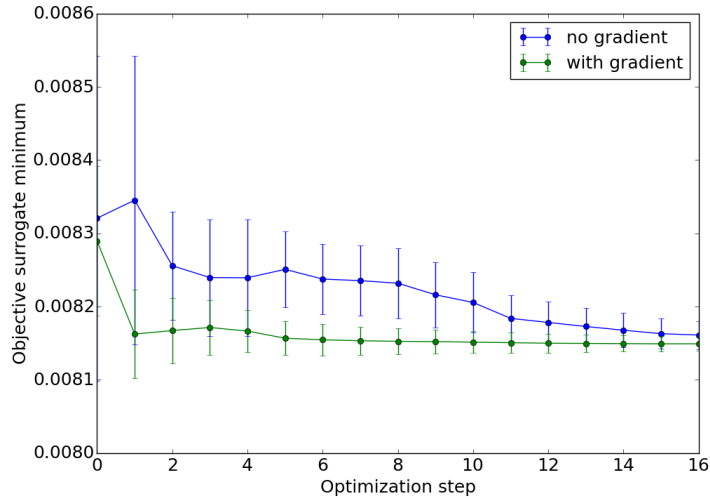}
\end{center}
\caption{Comparison between Bayesian optimization with and without gradients}
\label{f:grad_compare}
\end{figure}

In order to quantify the utility of the gradient in the optimization
process a comparison is performed between Bayesian optimization with and without gradients.
Running a separate optimization without gradients is too expensive as this
would involve running a large number of large eddy simulations. Hence,
the two Bayesian optimization algorithms are compared by averaging multiple optimization
runs that use an approximate LES design objective function.
This objective is a Gaussian process which is trained on all the objective
and gradient evaluations obtained during the adjoint-based design optimization process.
Both the Bayesian optimization
algorithms utilize the same DoE evaluations obtained during the adjoint-based
design optimization.
While running the optimizations, whenever the algorithm requires an evaluation  
for a particular design point, the objective is evaluated (including gradients if required) by sampling the
Gaussian process at that point using Equation \ref{e:first} by replacing $f(\bm{x})$ and
$f^\prime(\bm{x})$ with the mean of the corresponding Gaussian process. The noise term
is realized by sampling the normal distribution with zero mean 
and variance computed from the
mean of the noise log GP in accordance with Equation \ref{e:second}.

Figure \ref{f:grad_compare} shows the 
trajectories of the minimum objective value for the two optimization algorithms
as a function of the number of function evaluations.
The minimum objective value as reported by the optimizers is
minimum of the mean function of the posterior GP of the optimization algorithm.
The figure shows a trajectory of the mean and the uncertainty corresponding to a single standard deviation of the
posterior GP at the design point with the minimum objective value. 
The Bayesian optimizers with and 
without gradients are executed $1000$ times to achieve a statistically converged Monte Carlo estimate of
the optimization trajectory. The estimates provide less than $1\%$ relative sample standard deviation 
for each mean value in the optimization trajectory.

From the figure it can be seen that the Bayesian optimization with gradients performs significantly
better than optimizer without gradients as it reaches the optimal design faster.
Furthermore, the optimizer with gradients obtains the minimum design objective
value with a much lower standard deviation.

\section{Conclusion}
\label{s:conclusion}
Adjoint-based design optimization using large eddy simulations is a powerful tool
for design of turbomachinery components. Using a viscosity-stabilized adjoint
method and a gradient utilizing Bayesian optimization method, the trailing
edge of a gas turbine vane was optimized. The optimal design has a 
$12\%$ reduction in Nusselt number and $16\%$ reduction in pressure loss coefficient.

The Bayesian optimizer has a few issues scaling to a large number of design 
parameters ($> 20$). Further research needs to be performed in order to either 
increase it's efficiency for higher dimensional problems or investigate alternative optimization algorithms like multi-start 
stochastic gradient descent. 
The success of the adjoint-based design 
optimization method using LES increases it's applicability
to more challenging design problems like shape design of
an entire low-pressure turbine blade.

\section{Acknowledgements}
The authors thank GE Aviation for providing funding and valuable feedback for this project. 
This paper uses results obtained from
the Mira supercomputer of the Argonne Leadership Computing Facility, 
which is a DOE Office of Science User Facility supported under Contract DE-AC02-06CH11357.

\bibliographystyle{asmems4}


%
% ...

\bibliography{optim}

\end{document}